\documentclass[twocolumn]{aa}
\usepackage{graphicx}
\usepackage{psfig}
\def\ltsima{$\; \buildrel < \over \sim \;$}
\def\simlt{\lower.5ex\hbox{\ltsima}}
\def\gtsima{$\; \buildrel > \over \sim \;$}
\def\simgt{\lower.5ex\hbox{\gtsima}}
\begin{document}

   \title{Probing the formation of intermediate- to high-mass 
stars in protoclusters}

   \subtitle{A detailed millimeter study of the NGC~2264 clumps}
\titlerunning{Millimeter study of NGC~2264}
   \author{N. Peretto
          \inst{1}, 
          Ph. Andr\'e\inst{1,2}, 
          \and 
	  A. Belloche\inst{3} 
          }

   \offprints{N. Peretto or Ph. Andr\'e}

   \institute{$^1$Service d'Astrophysique, CEA/DSM/DAPNIA, C.E. Saclay,
              Orme des Merisiers, 91191 Gif-sur-Yvette, France\\
	      $^2$AIM -- Unité Mixte de Recherche CEA -- CNRS 
		  -- Université Paris VII -- UMR n° 7158\\
	      $^3$Max Planck Institut f\"ur Radioastronomie, 
Auf dem H\"ugel 69, 53121 Bonn, Germany\\
              \email{peretto@cea.fr, pandre@cea.fr}
             }

   \date{Received April 27, 2005}

   \abstract{
We present the results of dust continuum and molecular line observations 
of two massive cluster-forming clumps, NGC~2264-C and NGC~2264-D, including
extensive mapping performed with the MAMBO bolometer array and the HERA 
heterodyne array on the IRAM 30m telescope. Both NGC~2264 clumps are located in 
the Mon OB1 giant molecular cloud complex, adjacent to one another.
Twelve and fifteen compact millimeter continuum sources (i.e. MMSs) are
identified in clumps C and D, respectively. 
These MMSs have larger sizes and masses than the millimeter continuum 
condensations detected in well-known nearby protoclusters such as 
$\rho$~Ophiuchi.
The MMSs of NGC~2264 are more similar in size
to the DCO$^+$ `cores' of $\rho$~Oph, although they are somewhat 
denser and exhibit broader linewidths. Most of the MMSs of 
NGC~2264-C harbor candidate Class~0 protostars associated 
with shocked molecular hydrogen jets. 
Evidence for widespread infall motions is found in, e.g., 
HCO$^+$(3--2) or CS(3--2) in both NGC~2264-C and NGC~2264-D. 
A sharp velocity discontinuity $\sim$~2~km.s$^{-1}$ in amplitude is observed 
in N$_2$H$^+$(1--0) and H$^{13}$CO$^+$(1--0) in the central, innermost part 
of NGC~2264-C, which we interpret as the 
signature of a strong dynamical interaction between two MMSs and their 
possible merging with the central MMS C-MM3.
Radiative transfer modelling supports the idea that NGC~2264-C is 
a highly unstable prolate clump in the process of collapsing 
along its long axis on a near free-fall dynamical timescale 
$\sim$ 1.7$\times$10$^5$ yr.
Our model fit of this large-scale collapse suggests 
a maximum mass inflow rate 
$\sim$ 3$\times$10$^{-3}$ M$_{\odot}$~yr$^{-1}$ toward the central 
protostellar object C-MM3. 
In NGC~2264-D, we estimate a mass infall rate $\dot{M}_{DMM1} 
\sim$ 1.1$\times$10$^{-4}$ M$_{\odot}$ yr$^{-1}$ toward the rotating Class~0 
object D-MM1, also based on radiative transfer modelling of the 
observations.
Such infall rates are sufficiently high to overcome radiation pressure 
and allow the formation of $\sim 20$~M$_{\odot}$ stars by accretion 
in $\sim$ 1.7$\times$10$^5$~yr, i.e., a time similar 
to the global dynamical timescale of the central part of NGC~2264-C.
We conclude that we are likely witnessing the formation of a high-mass 
($\simgt 10$~M$_{\odot}$) protostar in the central part of NGC~2264-C.
Our results suggest a picture of massive star formation intermediate
between the scenario of stellar mergers of Bonnell et al.
(1998) and the massive turbulent core model of McKee \& Tan (2003), 
whereby a turbulent, massive ultra-dense core is formed 
by the gravitational merger of two or more Class~0 protostellar cores 
at the center of a collapsing protocluster.

\keywords{stars: formation -- stars: circumstellar matter -- stars: kinematics -- ISM: clouds -- ISM: structure -- ISM: kinematics and dynamics -- ISM: individual object: NGC~2264}
}

   \maketitle
%

\section{Introduction}

 While most stars are believed to form in clusters (e.g. Adams \& Myers 2001,
Lada \& Lada 2003), our present theoretical understanding of the star 
formation process is essentially limited to isolated dense cores and 
protostars (e.g. Shu et al. 1987, 2004). 
Detailed studies of the earliest phases of clustered star formation are 
crucially needed if we are to explain the origin of the stellar initial
mass function (IMF) and the birth of massive stars.
There is indeed a growing body of evidence that high-mass 
($M_\star > 8\, M_\odot $) stars may be able to form {\it only} in a 
clustered environment (e.g. Zinnecker et al. 1993, 
Testi et al. 1999, Clarke et al. 2000).

On the theoretical side, two main scenarios have been proposed 
for clustered star formation in turbulent molecular clouds.
In the first scenario, the distribution of stellar masses 
is primarily determined by {\it turbulent fragmentation}.
Briefly, self-gravitating 
condensations (each containing one local Jeans
mass) form as turbulence-generated density fluctuations 
(e.g. Padoan \& Nordlund 2002), then decouple from 
their turbulent environment (e.g. Myers 1998), 
and eventually collapse with little interaction with their surroundings. 
In this scenario, high-mass stars are built-up by a scaled up 
version of the accretion process believed to be at work in low-mass 
protostars: a high accretion rate ($> 10^{-3}\, M_\odot$.yr$^{-1}$), generated 
by the high pressure of the turbulent environment (e.g. McKee \& Tan 2003)
and/or the influence of an external trigger (e.g. Hennebelle et al. 2003), 
is sufficient to overcome the radiation pressure that would normally 
halt the collapse soon after $M_\star \approx 8\, M_\odot $ 
(e.g. Stahler et al. 2000).\\
By contrast, in the second scenario, the distribution of stellar masses 
results entirely from the {\it dynamics of the parent protocluster}. 
Here, individual protostellar seeds have large proper motions with 
respect to the mean cloud velocity. They accrete cloud material competitively
while orbiting in the gravitational potential well of a collapsing 
protocluster and possibly collide with one another 
(e.g. Bonnell et al. 1998, 2001).  
In this alternative scenario, competitive accretion and 
dynamical interactions between individual cluster members play a dominant role 
(e.g. Klessen \& Burkert 2000, Bate et al. 2003). In the dense central 
region of the collapsing protocluster, the probability of encounters 
becomes large enough that massive stars can form by coalescence of 
intermediate-mass protostars (e.g. Bonnell \& Bate 2002).
Determining which of these two pictures dominates in actual cluster-forming 
clouds is a major open issue.

As one expects the internal dynamics of protoclusters to differ markedly in the
two pictures, comprehensive molecular line studies of the kinematics of
prestellar condensations and Class 0 protostars in cluster-forming regions 
can provide powerful observational tests (cf. Andr\'e 2002).
In an effort to improve our knowledge of the dynamics of protoclusters 
and constrain theoretical models of clustered star formation, we carried out a
combination of millimeter dust continuum and molecular line observations of 
the active cluster-forming region NGC~2264, 
containing more than 360 near-IR sources
(Lada et al. 1993, Lada \& Lada 2003). Located in the Mon OB1 molecular 
cloud complex (d$\sim$ 800 pc), NGC~2264 is known for its famous Cone Nebula. 
Three arcminutes or $\sim 0.7$~pc to the north of the Cone Nebula, 
Allen (1972) discovered a bright embedded IR source, hereafter called IRS1, 
associated with IRAS~06384+0932, and also known as Allen's source. 
IRS1 is a B2-type object with estimated visual extinction 
A$_{V}$ $\sim$ 20-30 (Allen 1972, Thompson et al. 1998),
bolometric luminosity $L_{bol} \sim 2300\, L_{\odot}$ (Margulis et al 1989), 
and mass $M_\star \sim$ 9.5 M$_{\odot}$ (Thompson et al. 1998). 
Approximatively 6 arcminutes or $\sim 1.4$~pc to the north-west of IRS1, 
there is another IRAS source, IRAS~06382+0939, hereafter called IRS2.
IRS2 is a Class~I young stellar object (YSO) 
with $L_{bol} \sim  150\, L_{\odot}$ (Margulis et al. 1989). 
Both IRS1 and IRS2 are associated with molecular outflows, named NGC~2264-C 
and NGC~2264-D, respectively (Margulis et al. 1988), and dense molecular 
clumps revealed by CS(2-1) (Wolf-Chase et al. 1995) and submillimeter 
dust continuum (Ward-Thompson et al. 2000; Williams \& Garland 2002, Wolf-Chase
et al. 2003) 
observations.
An H$_{2}$O maser was found in each clump (Genzel et al. 1977; Mendoza et al.
1990), and a 44-GHz methanol maser was found in NGC~2264-C (Haschick et al. 
1990). Methanol masers are known to be clear signposts of ongoing 
intermediate/high-mass star formation (Minier et al. 2001).\\ 
Previous kinematical studies of NGC~2264-C have revealed a complex 
velocity structure (Kr\"ugel et al. 1987). Evidence for large-scale collapse 
in the entire clump, disrupted by outflow motions on small scales, was 
reported by Williams \& Garland (2002). 
Triggered star formation, due to an expanding shell 
$\sim 0.12$~pc in diameter around IRS1, was suggested by Nakano et al. (2003) 
based on observations with the Nobeyama millimeter interferometer.
These previous studies had either 
lower angular resolution and/or were focused on a much smaller region
than the observations reported here.\\
The layout of the paper is as follows. Observational details,
dust continuum results, and molecular line results are described in 
Sect.~2, Sect.~3, and Sect.~4, respectively.  Section 5 discusses the nature 
of the embedded millimeter sources identified in NGC~2264, 
while Sect.~6 and Sect.~7 present radiative transfer models of NGC~2264-C 
and NGC~2264-D-MM1, respectively. We compare our observational results with
various scenarios of high-mass star formation in Sect.~8. 
Our main conclusions are summarized in Sect.~9. 


\section{Millimeter Observations}

\begin{figure*}
\psfig{file=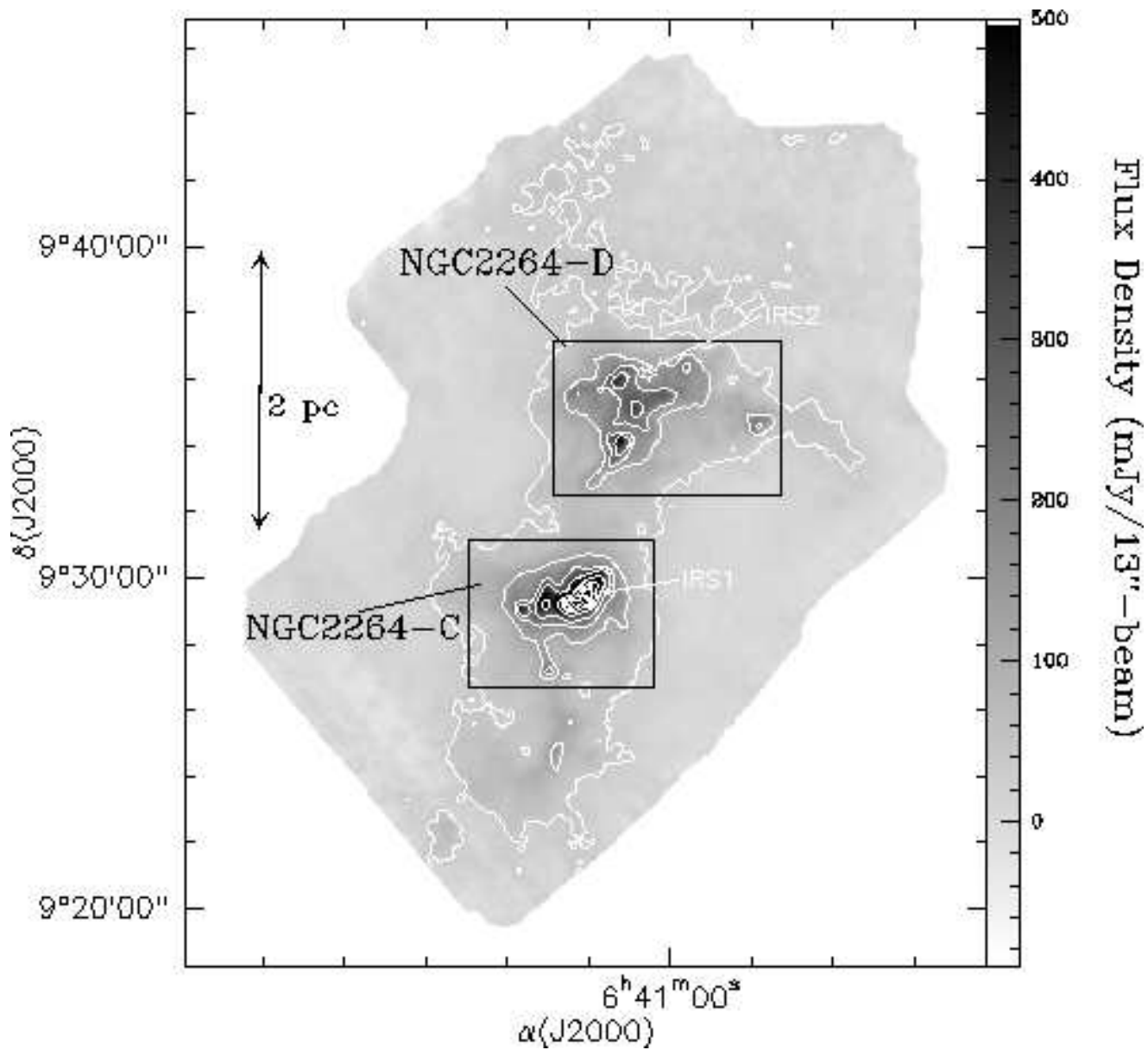,height=16cm,angle=0}
\caption{Millimeter dust continuum mosaic of NGC~2264-C and NGC~2264-D 
smoothed to an effective angular resolution of 13\arcsec .
The first white contour level is at 30 mJy/13\arcsec-beam and corresponds to a column
density $\sim$ 2$\times$10$^{22}$cm$^{-2}$ (assuming $\kappa$=0.005~cm$^{2}$.g$^{-1}$ and T$_d$=15~K). The other white contour levels go from 120 to 300 mJy/13\arcsec-beam by 90 mJy/13\arcsec-beam, and from 300 to 1100 mJy/13\arcsec-beam by 200 mJy/13\arcsec-beam. The mean rms noise is $\sim$ 8 mJy/13\arcsec-beam. The open white star symbols show the positions of the two IRAS sources IRS1 and IRS2.
\label{n2264tot.ps}}
\end{figure*}

We performed 1.2~mm dust continuum mapping observations of NGC~2264 
in December 2000 with the IRAM 30~m telescope near Pico Veleta, Spain, using 
the MAMBO 37-channel bolometer array. 
Eleven on-the-fly maps were taken in the dual-beam scanning mode, 
with individual sizes ranging from 13\arcmin $ \times$ 10\arcmin 
~to 5\arcmin$\times$4\arcmin. 
Chopping was performed by wobbling the secondary mirror of the telescope at 2~Hz 
with a throw ranging from 32\arcsec ~to 72\arcsec. The scanning speed was either 
6\arcsec/sec or 8\arcsec/sec. The atmospheric opacity at zenith
varied between 0.11 and 0.45 at $\lambda = 1.25$~mm. 
Pointing and focus checks were made every hour.  
The pointing accuracy was measured to be better than 4\arcsec ~and  
the angular resolution was $\sim$ 11\arcsec ~(HPBW). 
Calibration was achieved by mapping Mars (primary calibrator) and is believed
to be reliable to better than $\sim 15\% $.
Our resulting 1.2~mm continuum mosaic of the NGC~2264 region 
covers a total area of $\sim$ 400 
arcmin$^{2}$, i.e, $\sim$ 23 pc$^{2}$ with a mean rms noise $\sigma$ = 7 
mJy/11\arcsec-beam (i.e. 8 mJy/13\arcsec-beam, see Fig.~\ref{n2264tot.ps}).

We performed follow-up molecular line observations 
with the IRAM 30~m telescope in March and May 2002. 
The observed molecular transitions were 
N$_{2}$H$^+$(1-0), C$^{34}$S(2-1), HCO$^+$(1-0), H$^{13}$CO$^+$(1-0) at
3 mm, CS(3-2), C$^{34}$S(3-2) at 2 mm, CS(5-4), C$^{34}$S(5-4), HCO$^+$(3-2) 
and H$^{13}$CO$^+$(3-2)
at 1.1 mm. The half-power beam width of the telescope was $\sim$ 26\arcsec, 
$\sim$ 17\arcsec, $\sim$ 10\arcsec  at 3 mm, 2 mm, 1.1 mm, respectively. 
For most transitions, we took maps in the on-the-fly mode around the 
peaks seen in the 1.2 mm dust continuum map of Fig.~\ref{n2264tot.ps}. 
We used either four SIS heterodyne receivers simultaneously or, 
for the 1.2 mm band, the HERA 9-pixel heterodyne array. As backend we used an 
autocorrelation spectrometer with a spectral resolution of 20~kHz at 3~mm, and 
40~kHz at 2~mm and 1.2~mm. The corresponding velocity resolution ranged from 
0.05 to 0.08 km.s$^{-1}$ per channel. 


\section{Dust continuum mapping results}

\subsection{Source extraction}

\begin{figure}[ht!]
\hspace{-0.5cm} 
\psfig{file=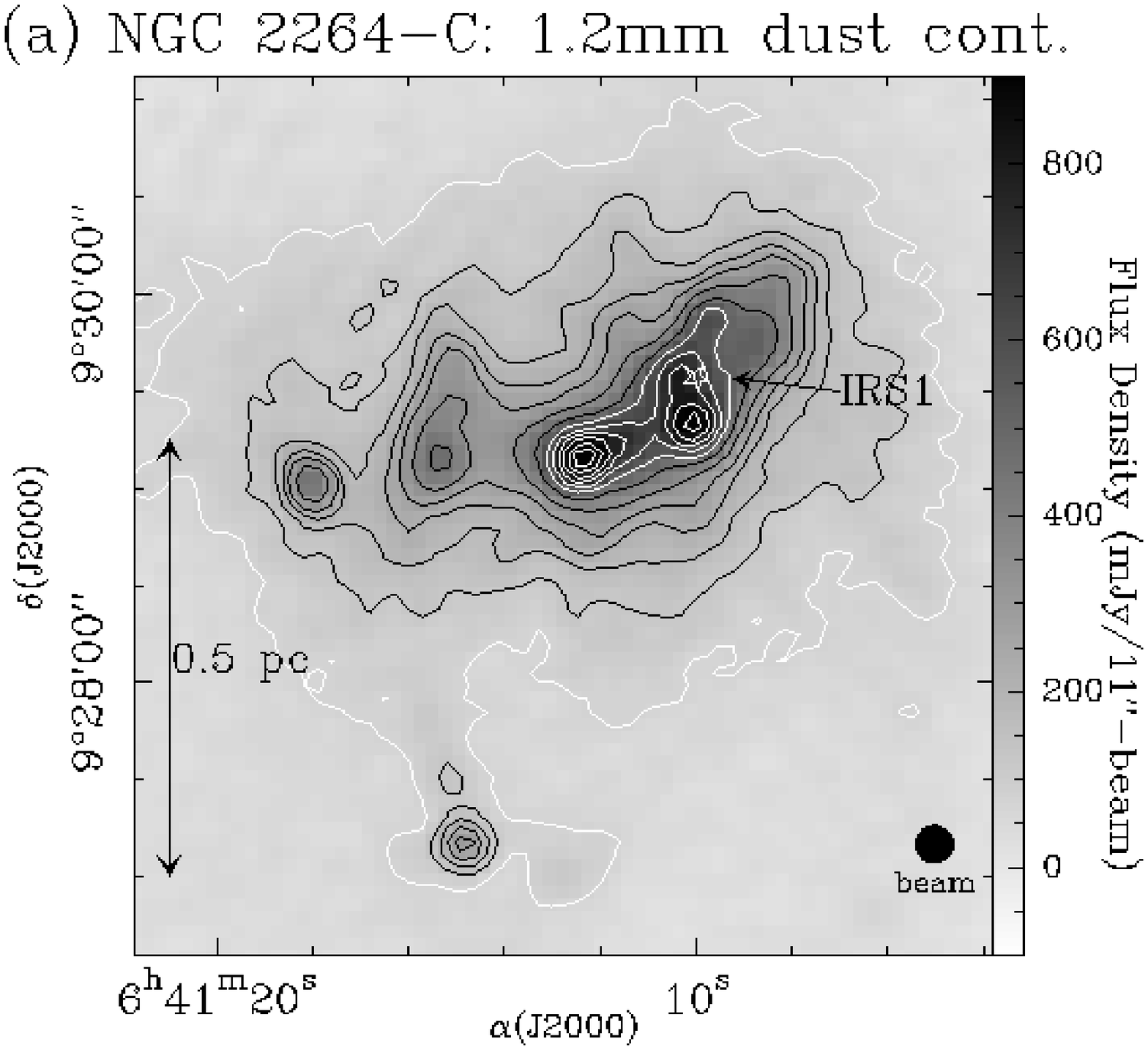,height=7cm,angle=0}

\hspace{-0.5cm}
\psfig{file=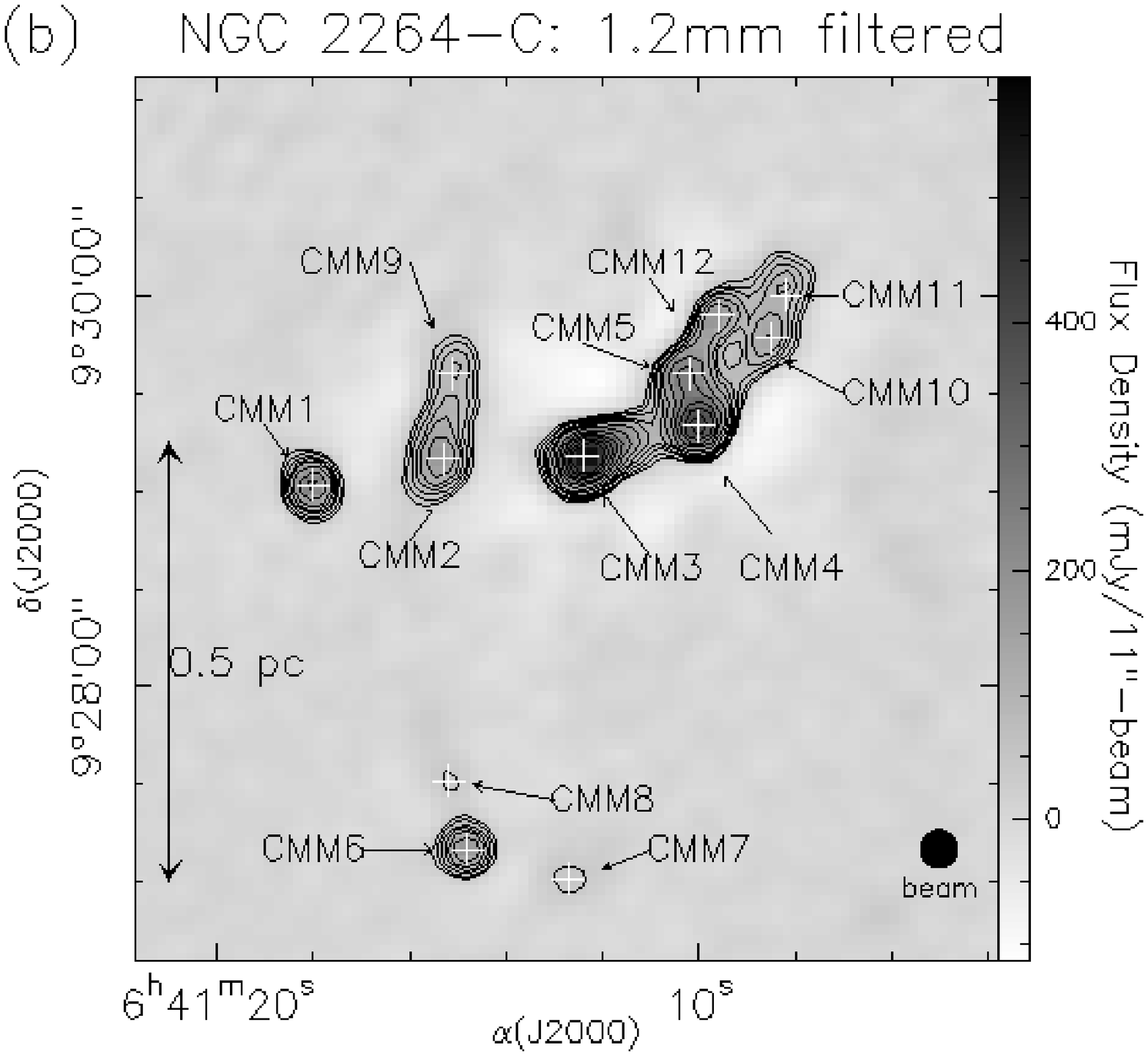,height=7cm,angle=0}
\caption{
\textbf{a)} Dust continuum map of NGC~2264-C at 1.2 mm.
The white open star shows the position of IRS1. The outer white contour 
corresponds
to S$_{peak}^{1.2}$=70 mJy/11\arcsec-beam (i.e. 10$\sigma$), equivalent to a 
column 
density $\sim$ 7$\times$10$^{22}$cm$^{-2}$ (assuming $\kappa$=0.005~cm$^{2}$.g$^{-1}$ and T$_d$=15~K), the black 
contours go from 120 to 270 by 50 mJy/11\arcsec-beam and from 350 to 450 by 100 mJy/11\arcsec-beam, while the inner white contours go from 550 to 1050 by 100mJy/11\arcsec-beam
\textbf{b)} Filtered 1.2mm dust continuum image of NGC~2264-C, obtained 
from the wavelet plane corresponding to spatial scales smaller 
than ScUp=24\arcsec. 
The white crosses mark the central positions of the MMSs identified 
with Gaussclump. The contours go from 35 to 50 by 15 mJy/11\arcsec-beam, from 50 to
110 by 20 mJy/11\arcsec-beam and from 140 to 440 by 50 mJy/11\arcsec-beam (See text for details.)
\label{n2264C.ps}}
\end{figure}
\begin{figure}[ht!]
\hspace{-0.8cm}
\psfig{file=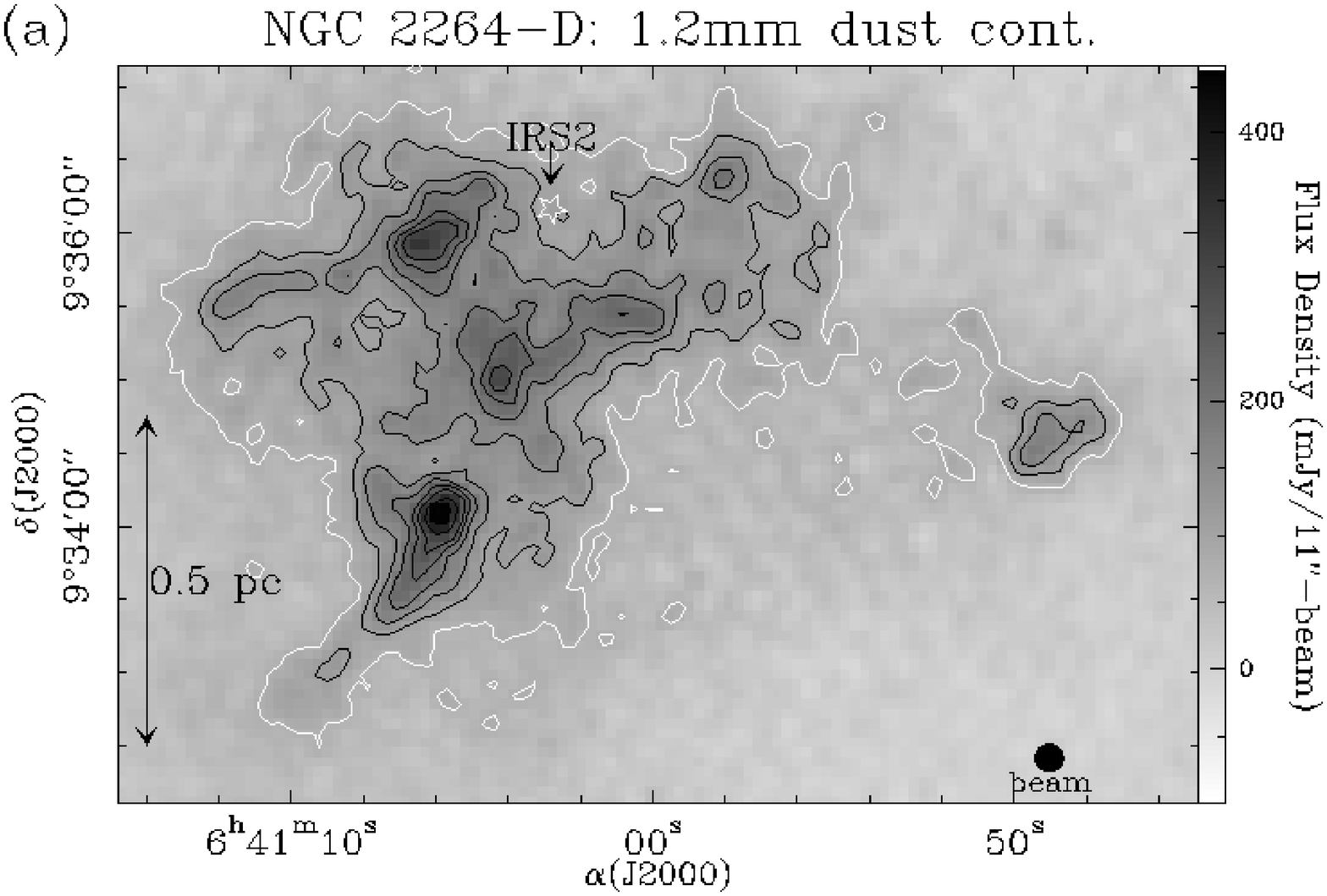,height=7cm,angle=0}

\hspace{-0.8cm}
\psfig{file=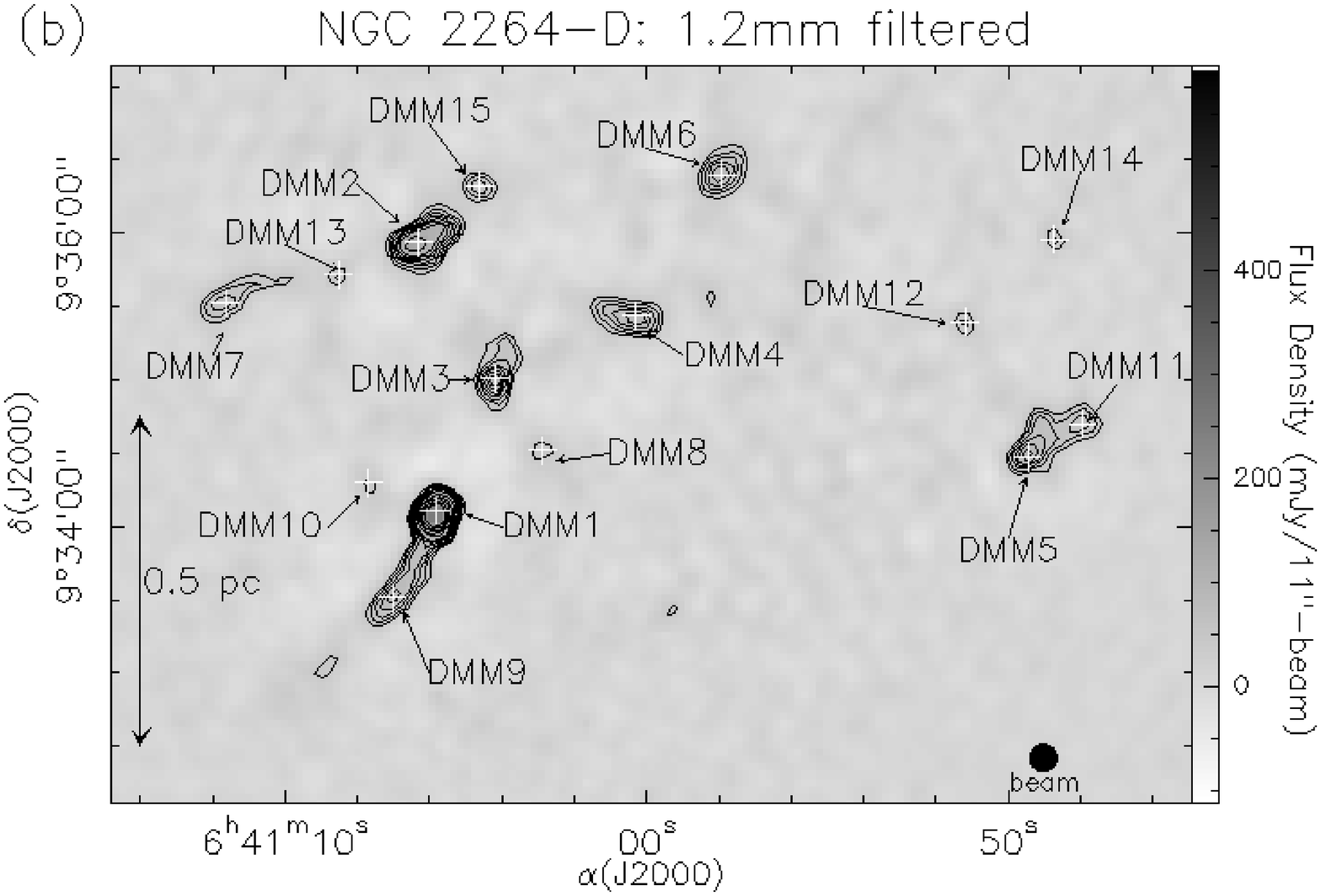,height=7cm,angle=0}
\caption{
Same as Fig.\ref{n2264C.ps} for NGC~2264-D. \textbf{a)} The white open star shows the position of IRS2. The outer white contour is the 
same as in Fig.~\ref{n2264C.ps}a, while the other black 
contours go from 110 to 270 by 40 mJy/11\arcsec-beam and from 350 to 425 by 75 mJy/11\arcsec-beam. \textbf{b)} The contours go from 35 to 95 by 10 mJy/11\arcsec-beam
 and from 120 to 200 by 20 mJy/11\arcsec-beam (See text for details.) 
\label{n2264D.ps}}
\end{figure}

Our 1.2 mm dust continuum maps were reduced and combined with the IRAM 
software for bolometer-array data (``NIC''; cf. Brogui\`ere et al. 1995). 
Combining all our maps we obtained the image shown in Fig.~\ref{n2264tot.ps}. 
The resulting mosaic confirms 
the presence of two distinct clumps, NGC~2264-C and NGC~2264-D, which closely 
follow the CS(2-1) integrated intensity map of Wolf-Chase et al. (1995).\\ 

\begin{table*}
\begin{minipage}[t]{\textwidth}
\caption{Source extraction results in NGC~2264-C and NGC~2264-D}        
\label{resume_C1}      
\centering                          
\renewcommand{\footnoterule}{}  
\begin{tabular}{c c c c c c c}        
\hline\hline                 
Source\footnote{The C-MM and D-MM numbers correspond to our own labelling, and are consistent with 
the source numbering of Ward-Thompson et al. (2000) and Wolf-Chase et al. (2003) for previously known sources. 
}      
& Coordinates & Undec.FWHM\footnote{Undeconvolved FWHM sizes derived from fitting an elliptical Gaussian to the background subtracted maps (i.e., after filtering out emission seen on scales larger than $ScUp =24\arcsec $)
} & P.A.\footnote{Position angle (from North to East) of the major axis of the fitted Gaussian ellipse } & S$_{peak}^{1.2}$\footnote{Peak flux 
density in the background subtracted continuum map and 1$\sigma$ rms at the position of the source}& $S_{int}^{1.2}$\footnote{Total integrated flux density under the elliptical Gaussian} & S$_{back}^{1.2}$\footnote{Background level at the position of the source peak}  \\
& $\alpha _{2000}$ $\,$ $\delta _{2000}$ & (arcsecond) & (deg)& (mJy/11\arcsec-beam)& (mJy)& (mJy/11\arcsec-beam)   \\    
\hline 
C-MM1     & 06:41:18.0 $\,$ +09:29:02 & 14$\times$12 & 16  & 255 $\pm$ 11 &354 & 186
\\
C-MM2     & 06:41:15.3 $\,$ +09:29:10 & 22$\times$13 & -12  & 183 $\pm$ 10 &433 & 296
\\
C-MM3    &06:41:12.4  $\,$ +09:29:11 & 18$\times$13 & 123  & 573 $\pm$ 9 &1108 & 508 
\\
C-MM4     & 06:41:10.0  $\,$ +09:29:20 & 18$\times$15 & -11  & 426 $\pm$ 8 &951 &541
\\
C-MM5    & 06:41:10.2  $\,$ +09:29:36 & 21$\times$11 &135  & 261 $\pm$ 9 &498 & 529 
\\
C-MM6     & 06:41:14.8 $\,$ +09:27:10 & 12$\times$12 & --  & 183 $\pm$ 6 & 218 & 100 
\\
C-MM7	  & 06:41:12.7  $\,$ +09:27:01 & 15$\times$13 & 126 & 45 $\pm$ 6 & 73 & 68 
\\
C-MM8	  & 06:41:15.2  $\,$ +09:27:31 & 14$\times$11 & 23  & 35 $\pm$ 6 & 45 & 89
\\
C-MM9     & 06:41:15.1  $\,$ +09:29:36 & 21$\times$11 & -1  & 94 $\pm$ 10 & 179 & 225
\\
C-MM10     & 06:41:08.5  $\,$ +09:29:47 & 27$\times$12 & -24 & 140 $\pm$ 8 &375 & 366
\\
C-MM11     & 06:41:08.2 $\,$ +09:30:00 & 16$\times$11 & 103  & 110 $\pm$ 7 &160& 288
\\
C-MM12   & 06:41:09.8  $\,$ +09:29:51 & 14$\times$11 & 121& 183 $\pm$ 7 & 233 &
387 
\\                    
\hline 
\hline
D-MM1     & 06:41:05.8  $\,$ +09:34:06 & 17$\times$13 & -24& 257 $\pm$ 8 & 469 & 211 
\\
D-MM2     & 06:41:06.3  $\,$ +09:35:56 & 23$\times$14 & 116 & 134 $\pm$ 9 &357 & 204 
\\
D-MM3    & 06:41:04.2  $\,$ +09:35:01 & 18$\times$12 & -11 &  105 $\pm$ 9 &187 & 201 
\\
D-MM4	  & 06:41:00.3  $\,$ +09:35:26 & 26$\times$13 & 81  & 81 $\pm$ 9 & 226 & 159
\\
D-MM5	  & 06:40:49.4  $\,$ +09:34:29 & 19$\times$13 & 135 & 88 $\pm$ 8 & 180 & 93
\\
D-MM6	  & 06:40:57.9  $\,$ +09:36:24 & 20$\times$15 & -39& 86 $\pm$ 9 & 213 & 119 
\\
D-MM7   & 06:41:11.6  $\,$ +09:35:32 & 28$\times$13 & 126 & 60 $\pm$ 8  & 180 & 116
\\
D-MM8     & 06:41:02.9  $\,$ +09:34:31 & 15$\times$11 & 33  & 41 $\pm$ 9 & 56 & 136
\\
D-MM9	  & 06:41:07.0  $\,$ +09:33:31 & 31$\times$12 & 139 & 77 $\pm$ 7 & 237 & 137
\\
D-MM10    & 06:41:07.7  $\,$ +09:34:18 & 16$\times$11 & 12& 36 $\pm$ 8 &  52 & 136
\\
D-MM11    & 06:40:47.9  $\,$ +09:34:42 & 19$\times$15 & 109 & 62 $\pm$ 8 & 146& 90 
\\
D-MM12	  & 06:40:51.2  $\,$ +09:35:24 & 12$\times$11 & 51 & 49 $\pm$ 10 & 53 & 53
\\
D-MM13    & 06:41:08.5  $\,$ +09:35:43 & 12$\times$11 & -34 & 46 $\pm$ 8 & 50 & 137
\\
D-MM14    &06:40:48.7   $\,$ +09:35:57 & 12$\times$11 & 29 & 46 $\pm$ 10 & 50& 26 
\\
D-MM15	  & 06:41:04.6  $\,$ +09:36:19 & 16$\times$12 &  102 & 62 $\pm$ 9 & 98 & 146
\\
\hline
\end{tabular}
\end{minipage}                           
\end{table*}

In order to search for compact millimeter continuum sources (hereafter, MMSs)
within NGC~2264-C and NGC~2264-D, we used the systematic method developed
by Motte, Schilke, \& Lis (2003). This method is  
based on a combination of the multi-resolution wavelet algorithm 
of Starck et al. (1995) and the Gaussian fitting procedure of 
Stutzki \& G\"usten (1990), Gaussclump (kindly provided by C.~Kramer). 
The original image is first decomposed into two wavelet 
``views'' of the field on small and large spatial scales, respectively.
The difficult step at this stage is to select a proper limiting scale, 
called ScUp, beyond which the detected dust emission is deemed to arise from 
the ambient cloud rather than from prestellar and/or protostellar sources.  
Here, we choose ScUp by analogy with the previous detailed study of the 
nearby $\rho$~Ophiuchi protocluster by Motte, Andr\'e, \& Neri (1998 -- 
hereafter MAN98). In $\rho$~Ophiuchi, the physical diameter of dense cores
(such as, e.g., Oph-A) was found to be $\sim$~0.1~pc, corresponding 
to an angular diameter $\sim$~24\arcsec ~at 800 pc. As the mean angular 
separation between nearest neighbouring peaks in our dust continuum 
map of NGC~2264-C is also $\sim$~24\arcsec, 
we adopt ScUp = 24\arcsec ~in NGC~2264-C. 
In order to facilitate the comparison of source masses at the same spatial 
scale in the entire NGC~2264 region, 
we adopt ScUp = 24\arcsec ~in NGC~2264-D as 
well, even though the mean angular separation between nearest neighbouring 
peaks is slightly larger ($\sim $~40\arcsec) in clump~D.
The wavelet decomposition then allows us to filter out all emission
structures arising from spatial scales larger than ScUp 
(see Fig.~\ref{n2264C.ps}b and Fig.~\ref{n2264D.ps}b).

In a second step, we use Gaussclump (Stutzki \& G\"usten 1990) to identify 
all Gaussian sources with FWHM sizes larger than the 11\arcsec
~beam width and above a given peak flux density threshold in the
filtered image (see Motte et al. 2003 for details).
Following this procedure with the detection threshold set to 
5$\sigma$ = 35 mJy/11\arcsec-beam, we identify  
a total of 27 MMSs~: 12 in NGC~2264-C (see Fig.~\ref{n2264C.ps}b) and 15 in 
NGC~2264-D (see Fig.~\ref{n2264D.ps}b), 14 of which are new 1.2 mm dust continuum 
detections. Table~\ref{resume_C1} lists the parameters of the 27 detected 
Gaussian MMSs. Although three more peaks lie just above 35~mJy/beam 
in the filtered map of NGC~2264-D (Fig.\ref{n2264D.ps}b),
they are not identified as real MMSs since their FWHM sizes are smaller than 
the beam width. 

\subsection{Properties of the NGC~2264 sources}

Assuming optically thin 1.2~mm dust continuum emission, the measured flux 
densities provide direct estimates of the masses and column densities of 
the MMSs (see, e.g., MAN98 for details).  We adopt a dust mass opacity 
$\kappa _{dust}$ = 0.005 cm$^{2}$.g$^{-1}$ at 1.2~mm (cf. MAN98) and 
a dust temperature $T_d = 15$~K (Ward-Thompson et al. 2000) for all sources. 
The distance to NGC~2264 is taken to be $d = 800$~pc and the mean molecular
weight $\mu = 2.33$. The derived masses, column densities, and volume densities
are listed in Table~\ref{table:4} for 
both NGC~2264-C and NGC~2264-D. 
Note that the masses of the compact MMSs account for only 
$\sim 10\% $ of the total gas mass of the clumps, the latter being
dominated by the large-scale background.

\begin{table*}
\begin{minipage}[t]{\textwidth}
\caption{ Properties of the millimeter sources detected in NGC~2264-C and NGC~2264-D} 
\label{table:4}      
\centering 
\renewcommand{\footnoterule}{}                         
\begin{tabular}{c c c c c c c c c}        
\hline\hline 
Source  & Dec.FWHM \footnote{Deconvolved major and minor FWHM continuum sizes; UR means unresolved} &  M$_{1.2}$\footnote{Mass estimated from the total 1.2~mm continuum flux density of the corresponding elliptical Gaussian source. For a 2D Gaussian source, $\sim 94\%$ of this mass is contained within an 
area twice the size of the FWHM ellipse. Typical uncertainty is a factor $\simgt 2$ 
(on either side) due to the uncertain values of the dust mass opacity and 
dust temperature} 
& Density\footnote{Mean H$_2$ volume density derived from the mass of col.~[3] assuming a spherical source 
of effective diameter twice the geometrical mean of the deconvolved FWHM sizes in col.~[2]. 
Typical uncertainty is a factor $\simgt 2$ as for M$_{1.2}$. The mean density within the FWHM volume is 
four times larger.} & 
N$_{H_2}$\footnote{Peak column density estimated from the background-subtracted map. 
Typical uncertainty is a factor $\simgt 2$ as for M$_{1.2}$} 
& N$_{H_2}^{back}$\footnote{Background column density at the position of the source peak 
assuming the same dust properties as for the MMSs (see text).}
& $\sigma _{line}$\footnote{Velocity dispersion (i.e. rms) along the line of sight derived from a Gaussian
hyperfine fit to the N$_2$H$^+$(1-0) multiplet. The typical fit uncertainty  
is $\Delta\sigma_{line}$=0.04 km.s$^{-1}$} & 
M$_{vir}$\footnote{Virial mass calculated as M$_{vir} =$ 3R$\frac{\sigma _{line}^2}{G}$ assuming a 
$\rho \propto$ r$^{-2}$ density profile. 
Here, the radius R is equal to twice the geometrical mean of the 
deconvolved HWHM radii. Typical relative uncertainty is less than 30$\%$} 
& $\alpha _{vir}$\footnote{Virial parameter defined as $\alpha _{vir}$ = M$_{vir}$/M$_{1.2}$.  
Typical uncertainty is a factor $\simgt 2$, dominated by the uncertainty on M$_{1.2}$ }\\
&  (pc) & (M$_{\odot}$) & (cm$^{-3}$) & (10$^{22}$cm$^{-2}$) & (10$^{22}$cm$^{-2}$) & (km.s$^{-1}$)& (M$_{\odot}$)& \\

\hline
C-MM1   & 0.035$\times$UR & 13.1 & 1.8$\times$10$^6$ & 26 & 19 & 0.38  & 3.2   & 0.2
\\
C-MM2   & 0.076$\times$0.028 & 16.0&  6.7$\times$10$^5$ & 18 & 30& 0.51 & 8.4 & 0.5
\\
C-MM3   & 0.057$\times$0.028 & 40.9 &  2.7$\times$10$^6$ & 57 & 51& 1.06& 31.3& 0.8
\\
C-MM4   & 0.057$\times$0.041 & 35.1 & 1.3$\times$10$^6$ & 43 & 54& 0.89 & 26.7& 0.8
\\
C-MM5   & 0.072$\times$UR & 18.4 &  8.4$\times$10$^5$ & 26 & 53& 0.93 & 27.1  & 1.5
\\
C-MM6   & unresolved  &  8.1  & $>$1.5$\times$10$^6$ & 18 & 10& 0.34 & 2.3   & 0.3
\\
C-MM7 & 0.041$\times$0.028 & 2.7 &  2.9$\times$10$^5$ & 5 & 7& 0.59 & 8.2   & 3.0
\\
C-MM8 & 0.035$\times$UR  & 1.7  &  2.3$\times$10$^5$ & 4 & 9& 0.47 & 4.8   & 2.8
\\
C-MM9  & 0.072$\times$UR  & 6.6  &  3.0$\times$10$^5$ & 9 & 23& 0.55 & 9.5   & 1.4
\\
C-MM10  & 0.099$\times$UR  & 13.9 &  3.9$\times$10$^5$ & 14 & 37& 0.76 & 21.2 & 1.5
\\
C-MM11  & 0.046$\times$UR &  5.9  &  5.3$\times$10$^5$ & 11 & 29& 0.68 & 11.6 & 2.0
\\
C-MM12 & 0.035$\times$UR & 8.6  &  1.2$\times$10$^6$ & 18 & 39& 0.76 & 12.6  & 1.5
\\
\hline
\hline
D-MM1  & 0.052$\times$0.028 & 17.3 &  1.3$\times$10$^6$ & 26 & 21& 0.59 & 9.3 & 0.5
\\
D-MM2  & 0.081$\times$0.035 & 13.2 &  3.6$\times$10$^5$ &  13 & 20& 0.51& 9.7 & 0.7
\\
D-MM3 & 0.057$\times$UR  &  6.9  &   4.5$\times$10$^5$ & 11 & 20& 0.26 & 1.9 & 0.3
\\
D-MM4 & 0.094$\times$0.028  &  8.4  &   2.6$\times$10$^5$ & 8 & 16& 0.81& 23.5& 2.8
\\
D-MM5 & 0.062$\times$0.028  &  6.7  &   3.8$\times$10$^5$ & 9 & 9& 0.42& 5.1 & 0.8
\\
D-MM6 & 0.067$\times$0.041  &  7.9  &   2.3$\times$10$^5$ & 9 & 12& 0.55& 11.0& 1.4
\\
D-MM7 & 0.103$\times$0.028  &  6.7  &   1.8$\times$10$^5$ & 6 & 12& 0.59& 13.0& 1.9
\\
D-MM8 & 0.041$\times$UR  & 2.1  &    2.2$\times$10$^5$ & 4 & 14& 0.34 & 2.7 & 1.3
\\
D-MM9 & 0.116$\times$UR  &  8.9  &    2.0$\times$10$^5$ & 8 & 14& 0.55 & 12.0 & 1.3
\\
D-MM10  & 0.046$\times$UR &  1.9  &   1.5$\times$10$^5$ & 4 & 14&0.72 & 13.0 & 6.8
\\
D-MM11 & 0.062$\times$0.041 &  5.4  &   1.7$\times$10$^5$ & 6 & 9& 0.30 & 3.2& 0.6
\\
D-MM12 & unresolved  & 1.9 &   $>$3.6$\times$10$^5$ & 5 & 5& 0.38 & 2.8  & 1.5
\\
D-MM13 & unresolved  &  1.8  &   $>$3.4$\times$10$^5$ & 5 & 14& 0.89 & 15.5 & 8.6
\\
D-MM14 & unresolved  &  1.8  &    $>$3.4$\times$10$^5$ & 5 & 3& -- & --  & --   
\\
D-MM15 & 0.046$\times$UR &  3.6  &   3.2$\times$10$^5$ & 6 & 15& 0.64 & 10.2 & 2.8
\\
\hline						
\end{tabular}
\end{minipage}
\end{table*}

Given the Gaussian fit uncertainties, a source is considered as unresolved 
if its undeconvolved FWHM angular size is smaller than 13\arcsec. 
Four sources are found to be unresolved (with both the major and minor 
FWHM sizes smaller than 13\arcsec), twelve are only partly resolved 
(with only the minor FWHM smaller than 13\arcsec), and eleven  
are fully resolved (with both the major and minor FWHMs larger than  
13\arcsec).
The MMSs have deconvolved FWHM physical sizes ranging from $<$ 0.03~pc to 0.12~pc, 
masses ranging from $\sim 2$ to $\sim 41$~M$_{\odot}$, and volume densities ranging from $\sim$ 1.5$\times$10$^5$ to  3$\times$10$^6$ cm$^{-3}$ (we can only estimate lower limits to the volume densities of the unresolved MMSs). 
Our 5$\sigma$ detection threshold corresponds to a mass of 
$\sim 1.2\, M_{\odot}$. 
A detailed comparison of the global properties of NGC~2264-C and NGC~2264-D 
is presented in \S~8 and Table~\ref{global} below. 
Here, we simply note that NGC~2264-C appears to be more centrally concentrated 
than NGC~2264-D in Fig.~1.

\section{Molecular line mapping results}

\subsection{N$_{2}$H$^{+}$ line widths and velocity dispersion}\label{vel_global} 

\begin{figure}[t!]
\hspace{0cm} 
\psfig{file=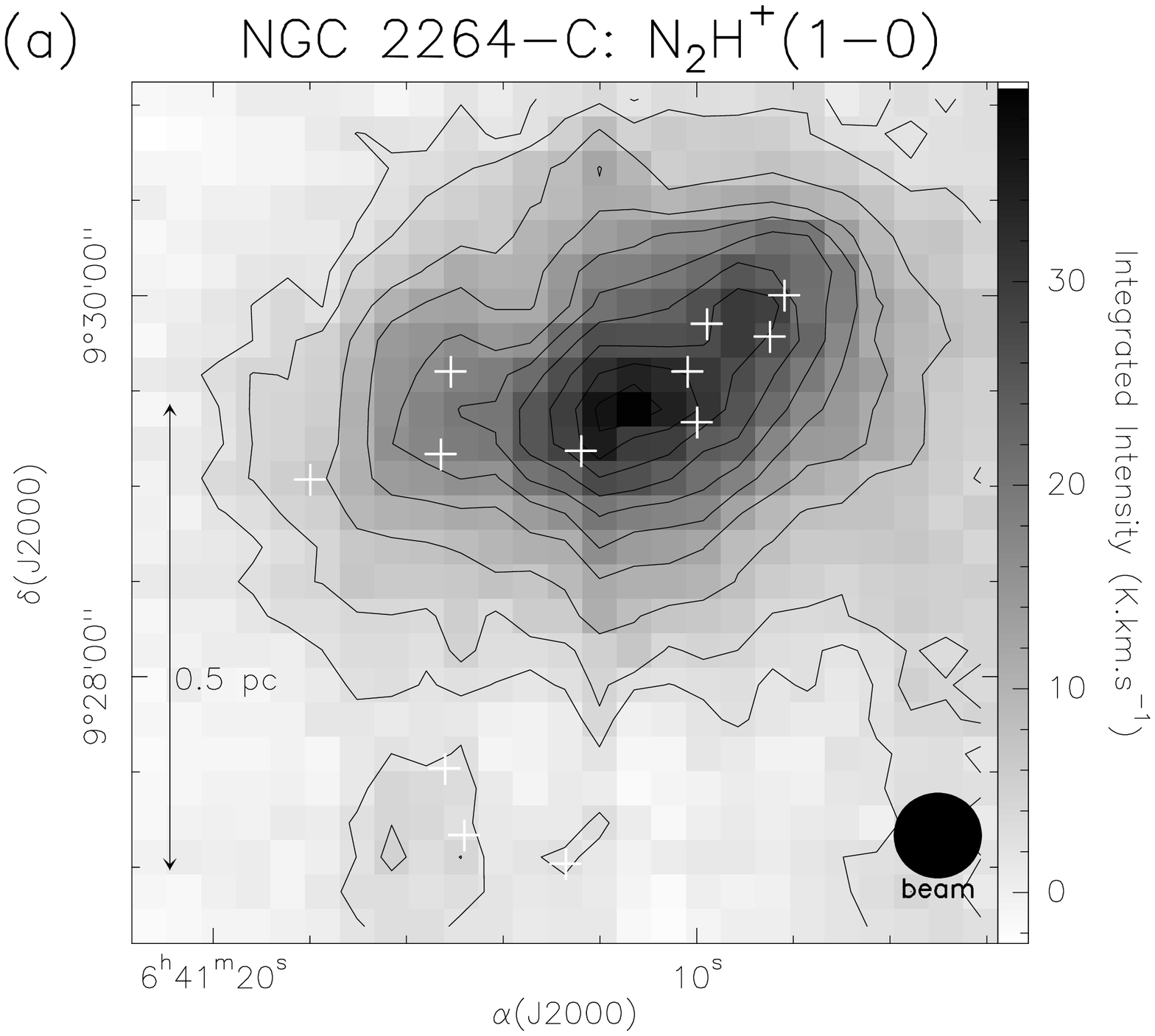,height=7cm,angle=270.}

\hspace{0.3cm}
\psfig{file=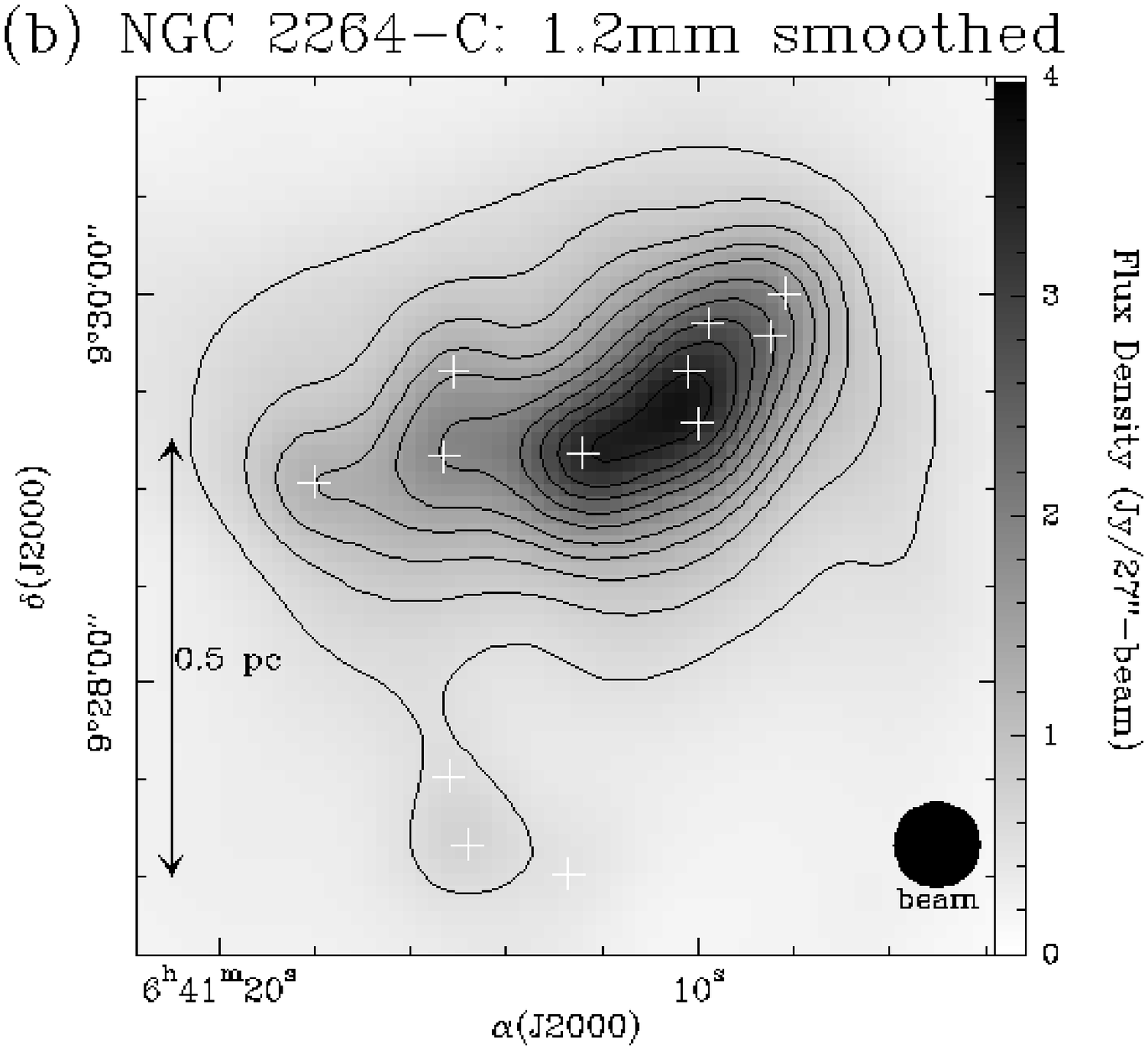,height=7cm,angle=0.}
\caption{
 \textbf{a)} Map of the N$_2$H$^+$(1-0) integrated line intensity from 0 to 25 
km.s$^{-1}$ in NGC~2264-C. First contour at 2 K.km.s$^{-1}$; other contours 
go from 4 to 36 K.km.s$^{-1}$ by 4 K.km.s$^{-1}$.
\textbf{b)}Millimeter dust continuum map of NGC~2264-C smoothed to the same 
27\arcsec ~(HPBW) angular resolution as the N$_2$H$^+$(1-0) map. Contour levels
go from 0.5 to 1.5 Jy/27\arcsec-beam by 0.25 Jy/27\arcsec-beam and from 1.5 to 3 Jy/27\arcsec-beam by 0.375 Jy/27\arcsec-beam. The white crosses mark the positions of the millimeter continuum sources identified in \S ~3. 
\label{n2264c_n2h+int.ps}}
\end{figure}

\begin{figure}[t!]
\hspace{-0.5cm}
\psfig{file=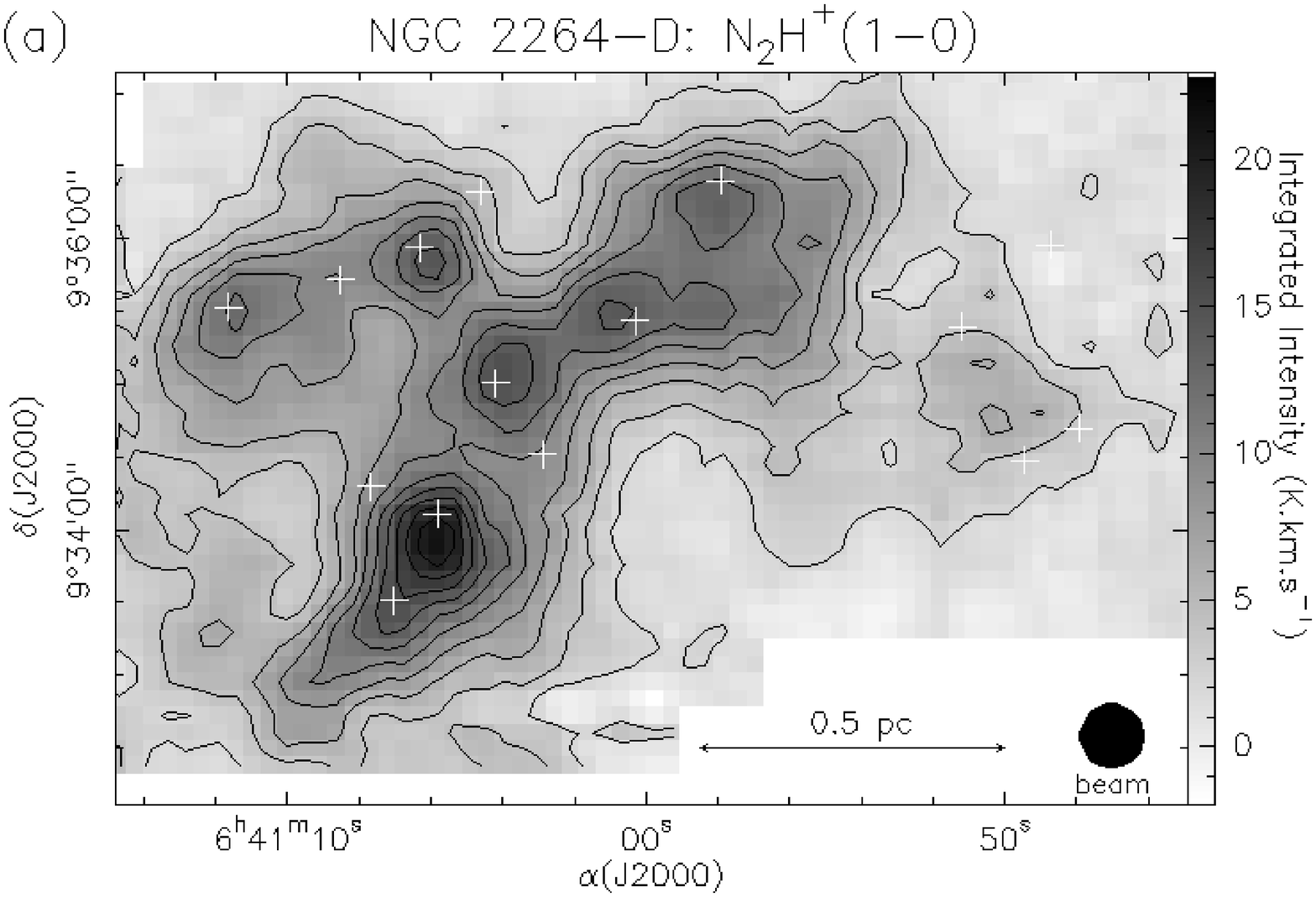,height=6.5cm,angle=0.}

\hspace{-0.5cm}
\psfig{file=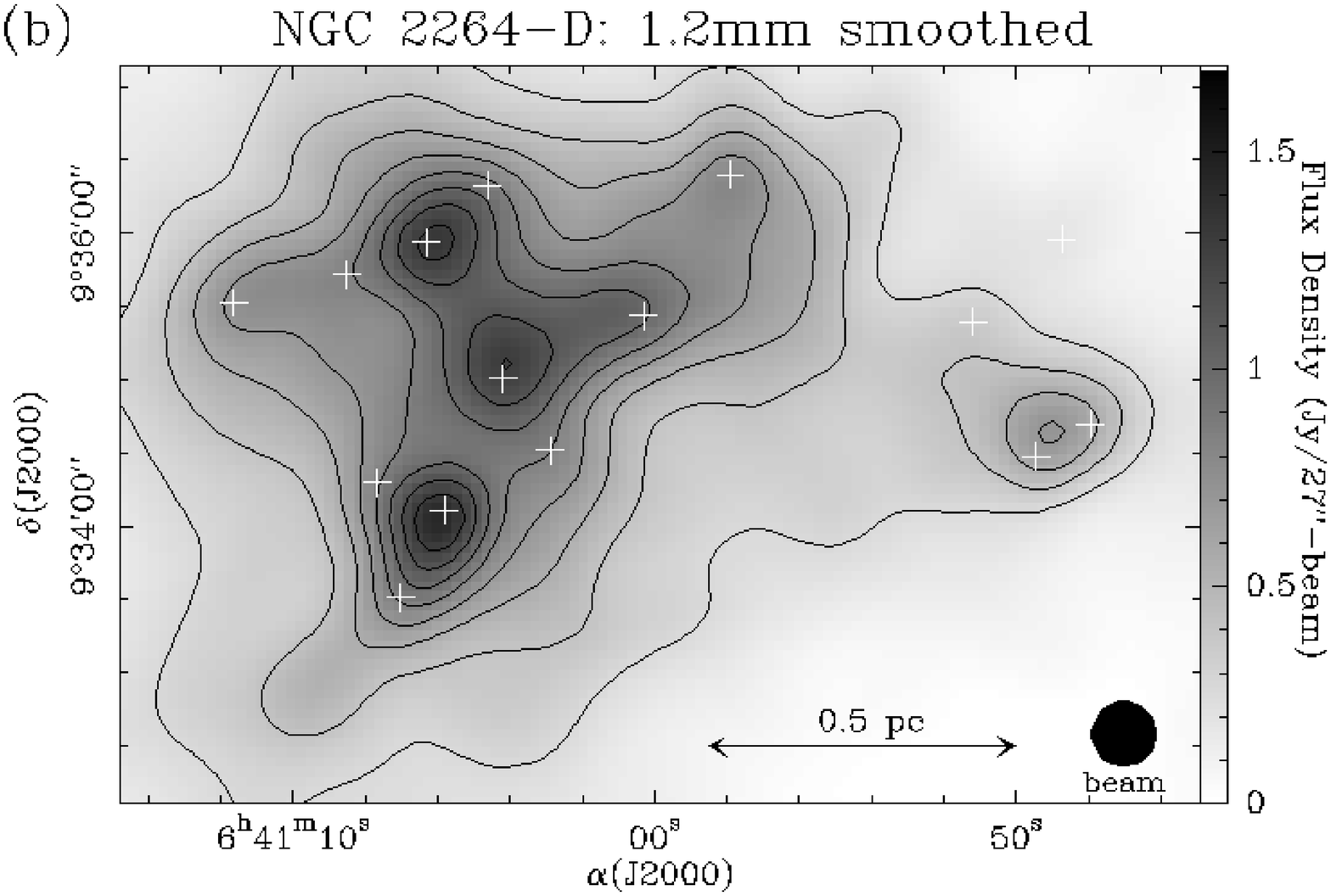,height=6.5cm,angle=0.}
\caption{
Same as Fig.\ref{n2264c_n2h+int.ps} for NGC~2264-D.
\textbf{a)} Contour levels go from 2 to 20 
K.km.s$^{-1}$ by 2 K.km.s$^{-1}$.
\textbf{b)} Contour levels go
from 0.25 to 1.3 Jy/27\arcsec-beam by 0.15 Jy/27\arcsec-beam.
\label{n2264d_n2h+int.ps}}
\end{figure}

Maps in low optical depth tracers of dense gas such as N$_{2}$H$^{+}$ 
can provide constraints on the velocity field of protoclusters projected 
onto the line of sight (cf. Belloche et al. 2001). 
The N$_{2}$H$^{+}$ molecule, which does not deplete 
up to fairly high densities (e.g. Bergin \& Langer 1997, Belloche \& 
Andr\'e 2004), is a particularly interesting tracer in this respect.
Figure~\ref{n2264c_n2h+int.ps} and Fig.~\ref{n2264d_n2h+int.ps} show that the N$_{2}$H$^{+}$(1-0) integrated
intensity maps of NGC~2264-C and NGC~2264-D trace essentially the same 
structures as the 1.2 mm dust continuum maps when the latter are smoothed 
to the same 27\arcsec  ~angular resolution.
  
We used the HFS (HyperFine Structure) fitting routine of the CLASS reduction 
package from IRAM to fit all 7 hyperfine components of the N$_2$H$^+$(1-0) 
multiplet at each mapped position. 
This provided, for each spectrum, estimates of the centroid velocity, 
V$_{LSR}$, and of the FWHM linewidth, $\Delta$V.


\begin{table}
\begin{minipage}[t]{\columnwidth}
\caption{Mean N$_2$H$^+$(1-0) line properties of clumps C and D}    
\label{resume_vel}      
\centering                          
\renewcommand{\footnoterule}{}
\begin{tabular}{c c c c c}        
\hline\hline 
Clump  & V$_{sys}$\footnote{Mean systemic velocity and rms dispersion around that value} 
&$<\sigma _{line}>$\footnote{Mean N$_2$H$^+$(1-0) line-of-sight velocity dispersion averaged over each clump and 
rms dispersion around that value}   & $\sigma _{1D}$\footnote{Standard deviation of the distribution of source centroid velocities measured in each clump. The error bar has been estimated 
as $\sqrt{\frac{2}{n-1}}\frac{\sigma_{1D}}{2}$, assuming that the source sample  
is drawn from a larger population whose velocity distribution follows Gaussian statistics. }   
& $\sigma _{3D}$\footnote{3D velocity dispersion of the MMSs calculated from
$\sigma _{1D}$ assuming isotropic motions. The error bar has been scaled from that 
estimated for $\sigma _{1D}$  }  \\
& (km.s$^{-1}$) & (km.s$^{-1}$) & (km.s$^{-1}$) & (km.s$^{-1}$)\\
\hline
NGC~2264-C &  7.5 $\pm$ 0.2  & 1.1 $\pm$ 0.1    & 0.7 $\pm$ 0.1  & 1.3 $\pm$ 0.3  \\
NGC~2264-D &  5.5 $\pm$ 0.2  & 1.0 $\pm$ 0.1    & 0.8 $\pm$ 0.2  & 1.4 $\pm$ 0.3 \\
\hline			
\end{tabular}
\end{minipage}
\end{table}

The N$_{2}$H$^{+}$ linewidth measured toward each MMS is listed  
as a velocity dispersion $\sigma _{line}$ = $\Delta V/\sqrt{8~ln(2)}$ 
in Table~\ref{table:4}.
On average, we find $\sigma _{line} \sim$ 0.6-0.7 km.s$^{-1}$ 
in both NGC~2264-C and NGC~2264-D.
From these $\sigma _{line}$ estimates, we can calculate a virial mass, 
and a virial mass ratio (cf. Bertoldi \& McKee 1992), 
$\alpha _{vir} = M_{vir}/M_{1.2}$, for each MMS 
(cf. Table~\ref{table:4}). 
Note that the N$_2$H$^+$(1-0) observations probe material on 27\arcsec ~scale, 
while the typical FWHM size of the dust continuum sources is only $\sim$ 15\arcsec.
However, since both M$_{vir}$ and M$_{1.2}$ are derived for a diameter $\sim$ twice the FWHM size,
our method of estimating $\alpha_{vir}$ should be reliable. 
Given the observational uncertainties on $M_{vir}$ and $M_{1.2}$, most MMSs are
consistent with virial equilibrium (i.e., $\alpha _{vir} \sim 1$). 
This is discussed further in \S~8. 
The measured values of $\sigma _{line}$ are to be compared with the thermal sound 
speed, c$_s$, which is $\sim$ 0.23 km.s$^{-1}$ for molecular gas at 15~K.
All MMSs have $\sigma _{line} > \rm{c}_s$, indicating that 
supersonic turbulence still dominates over thermal broadening down to spatial scales $\leq 0.1$~pc in the NGC~2264 region\footnote{In some cases, 
gravitational collapse may contribute to the linewidth and the value of 
M$_{vir}$ quoted in Table~\ref{table:4} may be correspondingly overestimated.
The radiative transfer models we present in \S~6 and \S~7 below suggest that 
infall contributes $\sim $~20-50\% of the linewidth for C-MM3 and D-MM1.}.
The typical Mach number $M = \sqrt{3} \frac{\sigma}{c_s}$ of the gas 
within the MMSs is larger than 5. 
On the larger $\sim 0.5-1 $~pc spatial scale of the NGC~2264
clumps, we estimate the Mach number to be $M \simgt 7$, based on the linewidth 
of the mean N$_{2}$H$^{+}$(1--0) spectrum averaged over the 
whole extent of NGC~2264-C and NGC~2264-D, respectively (cf. Table~\ref{resume_vel}). 
We conclude that NGC~2264 is characterized by supersonic motions down to the 
spatial scale of the MMSs.\\ 
We can also estimate the dispersion, $\sigma _{1D}$, of the line-of-sight 
velocities of the MMSs relatively to one another in each clump. 
Assuming isotropic motions, we can infer the 3D velocity dispersion of
the MMSs within each clump, $\sigma _{3D} = \sqrt{3}\,\sigma _{1D} $ 
(see Table~\ref{resume_vel}).  
Based on these velocity dispersion estimates, the crossing times of the MMSs 
across their parent clumps are calculated to be t$_{cross} \sim$ 6.0 $\times$ 10$^5$ yr in Clump C and $\sim$ 6.3 $\times$ 10$^5$ yr in Clump D.

\begin{figure*}[!ht]
\centerline{\hbox{
\hspace{-0.2cm} 
\psfig{file=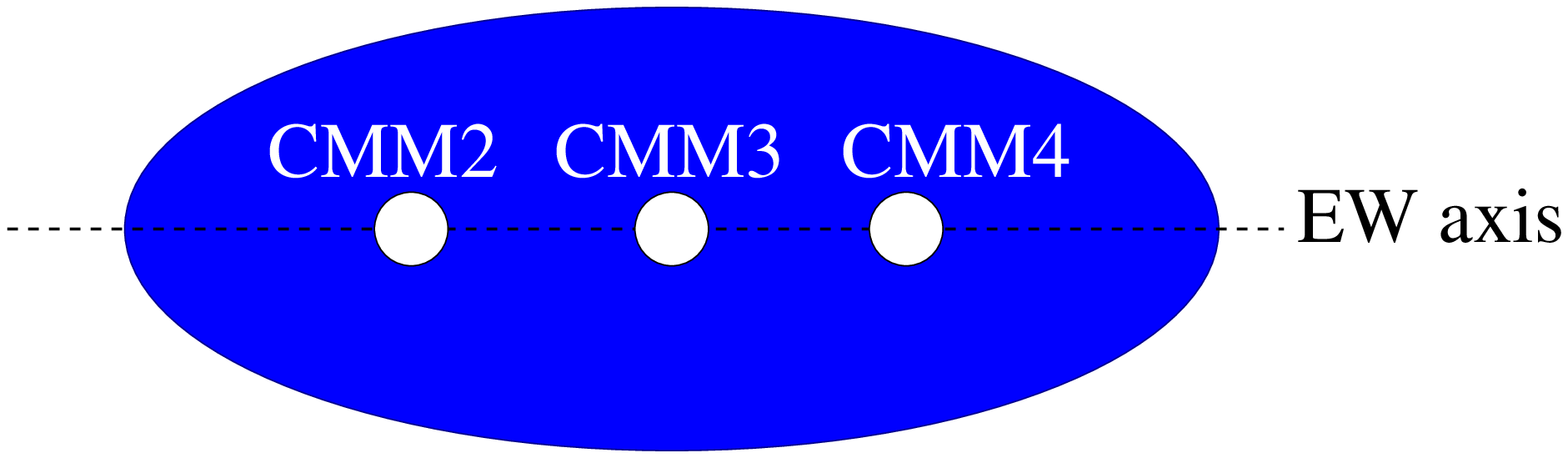,height=5.3cm,angle=270}
\hspace{-0.9cm}
\psfig{file=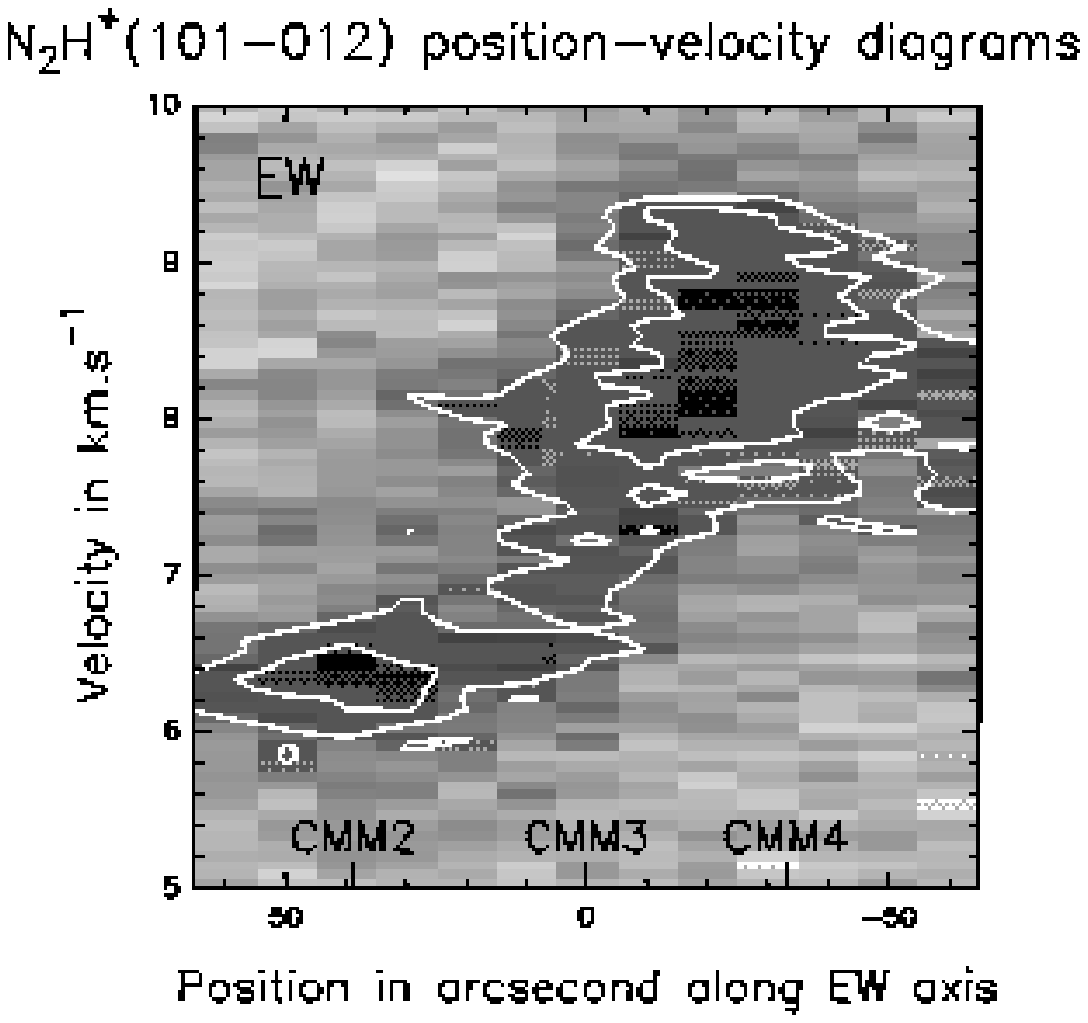,height=5.4cm,angle=0}
\hspace{-0cm}
\psfig{file=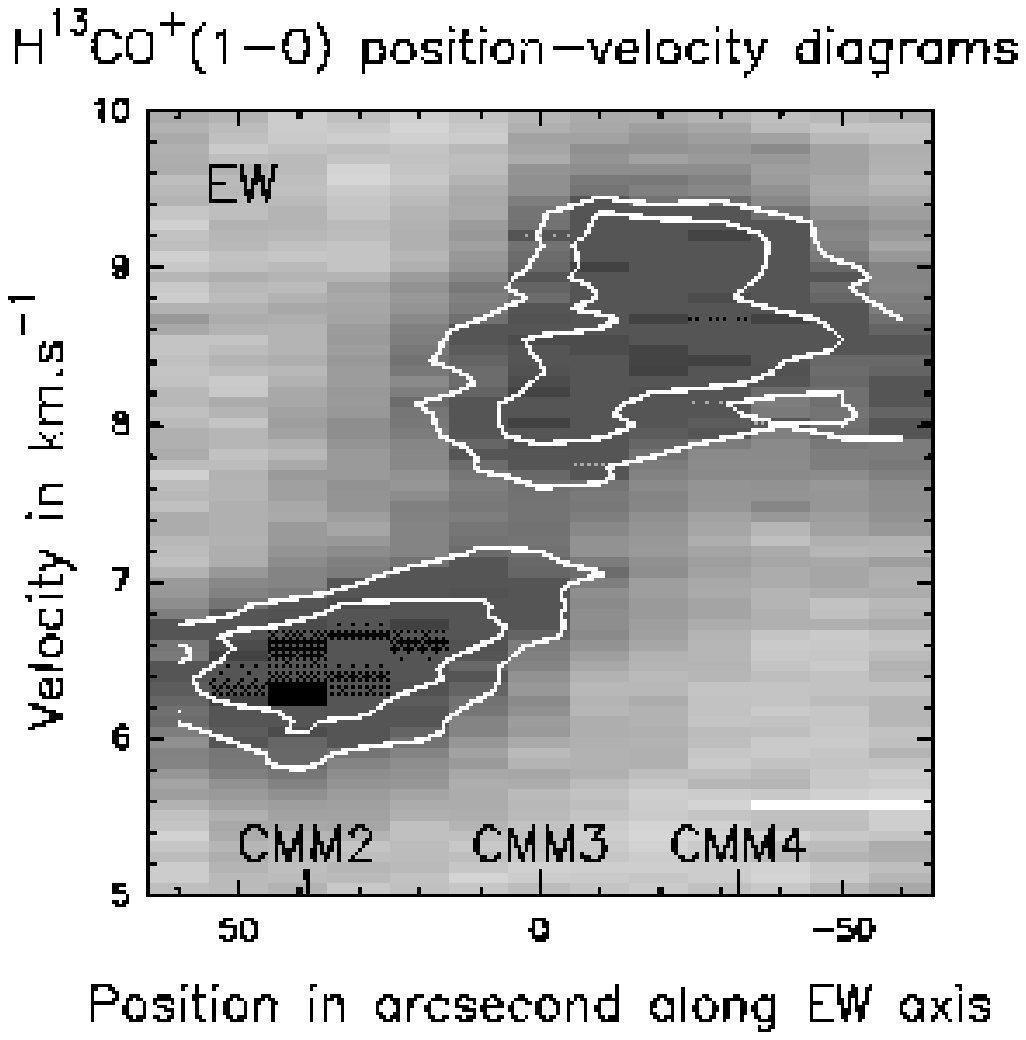,height=5.4cm,angle=0}}}
\vspace{-0.0cm}
\centerline{\hbox{
\hspace{-.5cm} 
\psfig{file=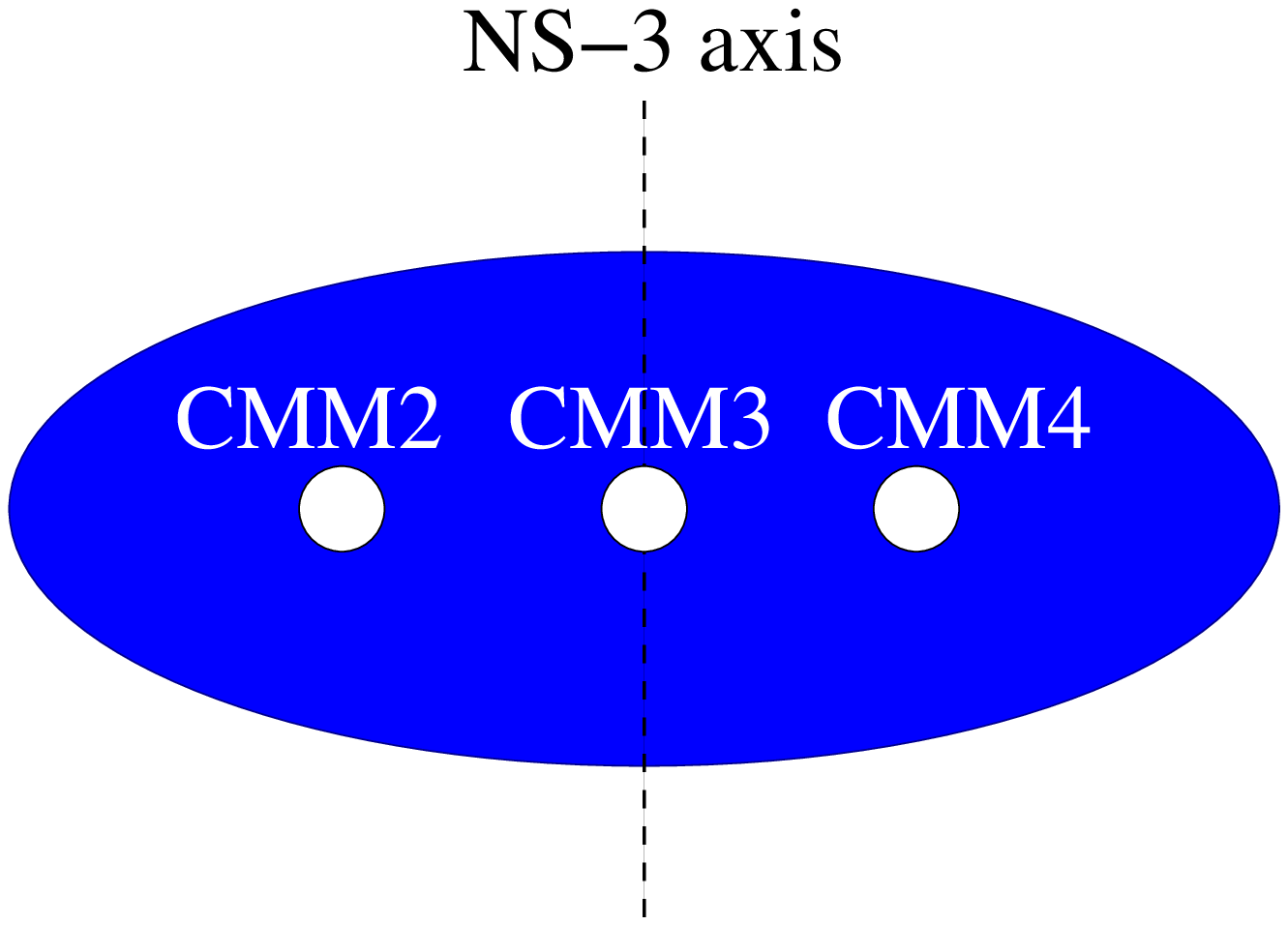,height=5.3cm,angle=270}
\hspace{-0.5cm}
\psfig{file=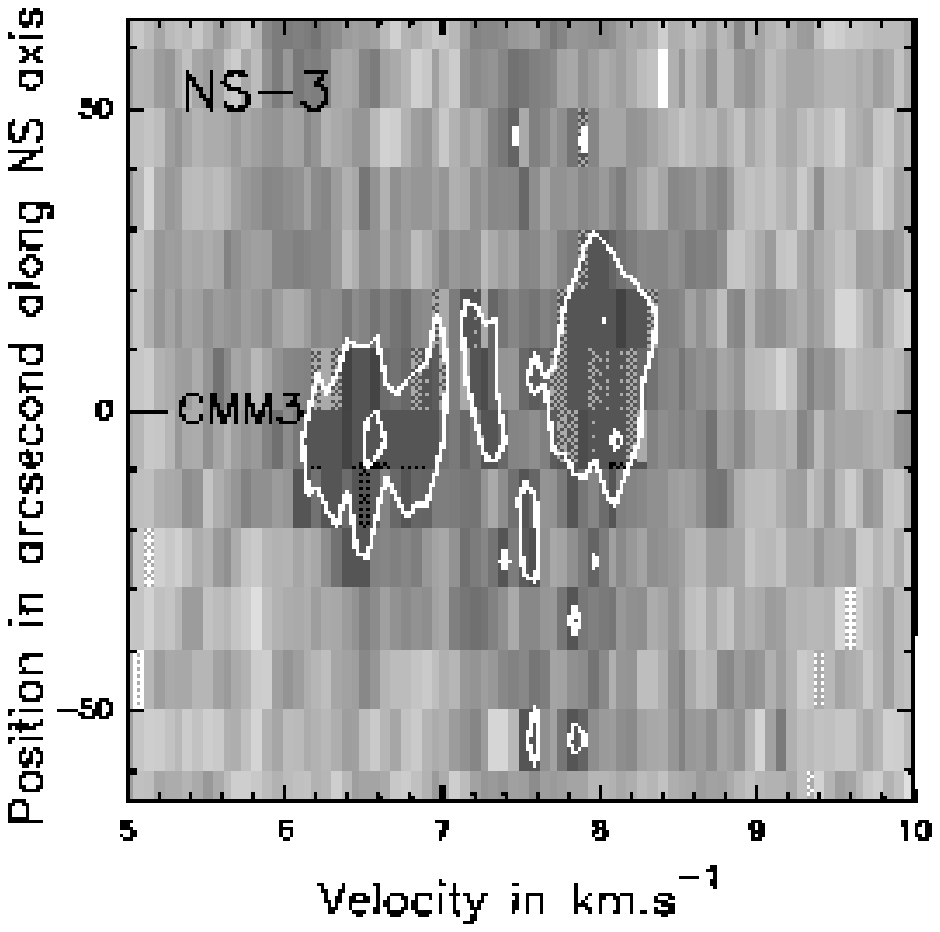,height=5.4cm,angle=0}
\hspace{-0cm}
\psfig{file=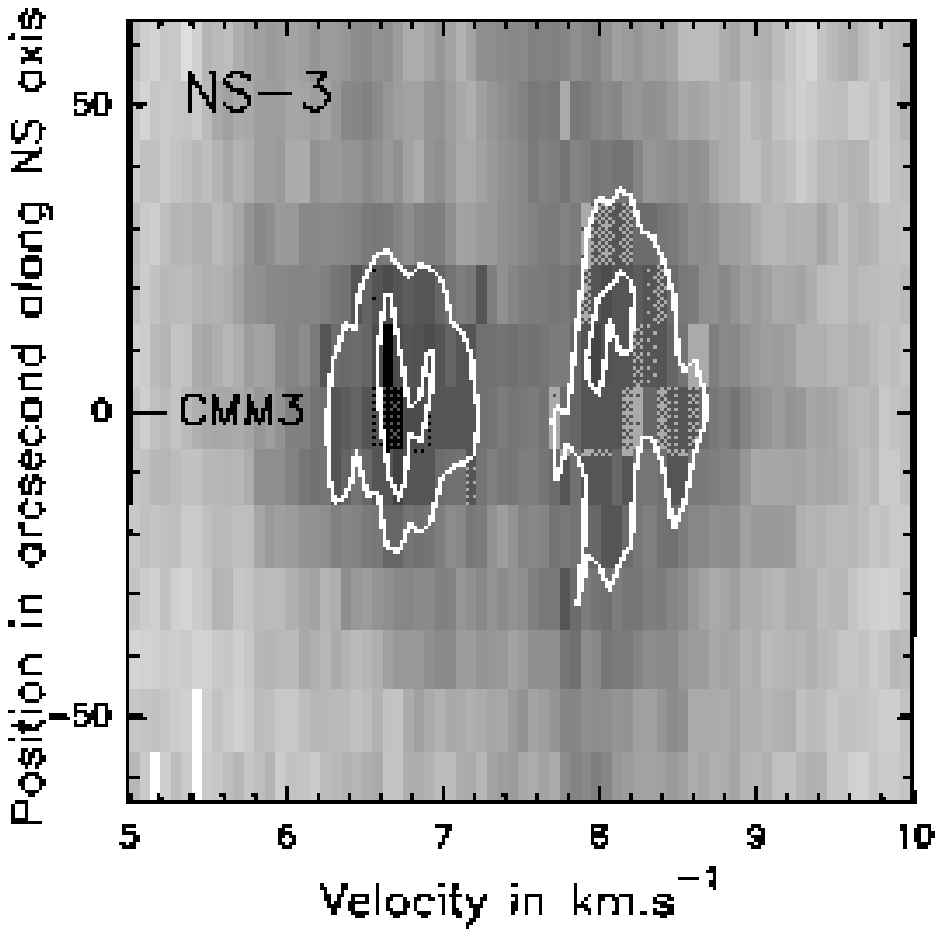,height=5.4cm,angle=0}}}
\vspace{-0.0cm}
\centerline{\hbox{
\hspace{-.5cm} 
\psfig{file=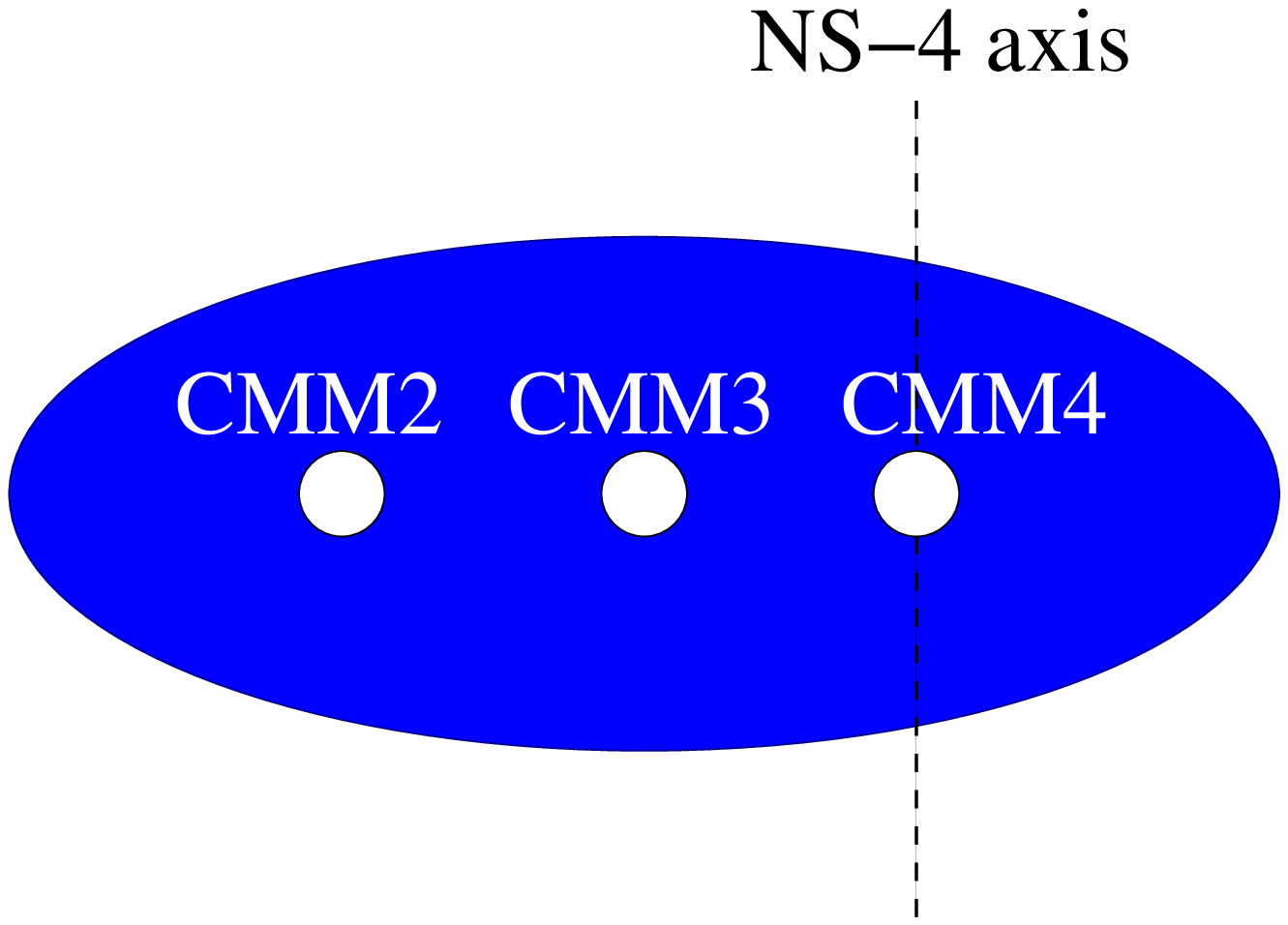,height=5.3cm,angle=270}
\hspace{-0.5cm}
\psfig{file=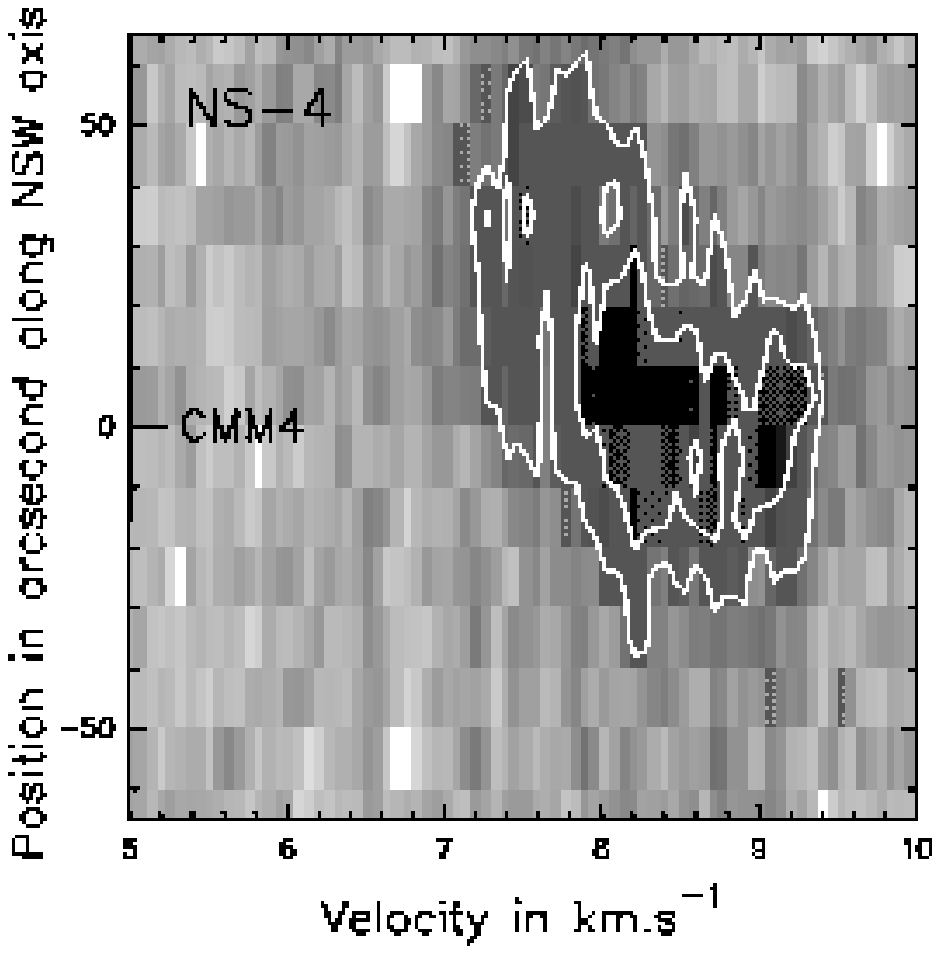,height=5.4cm,angle=0}
\hspace{-0cm}
\psfig{file=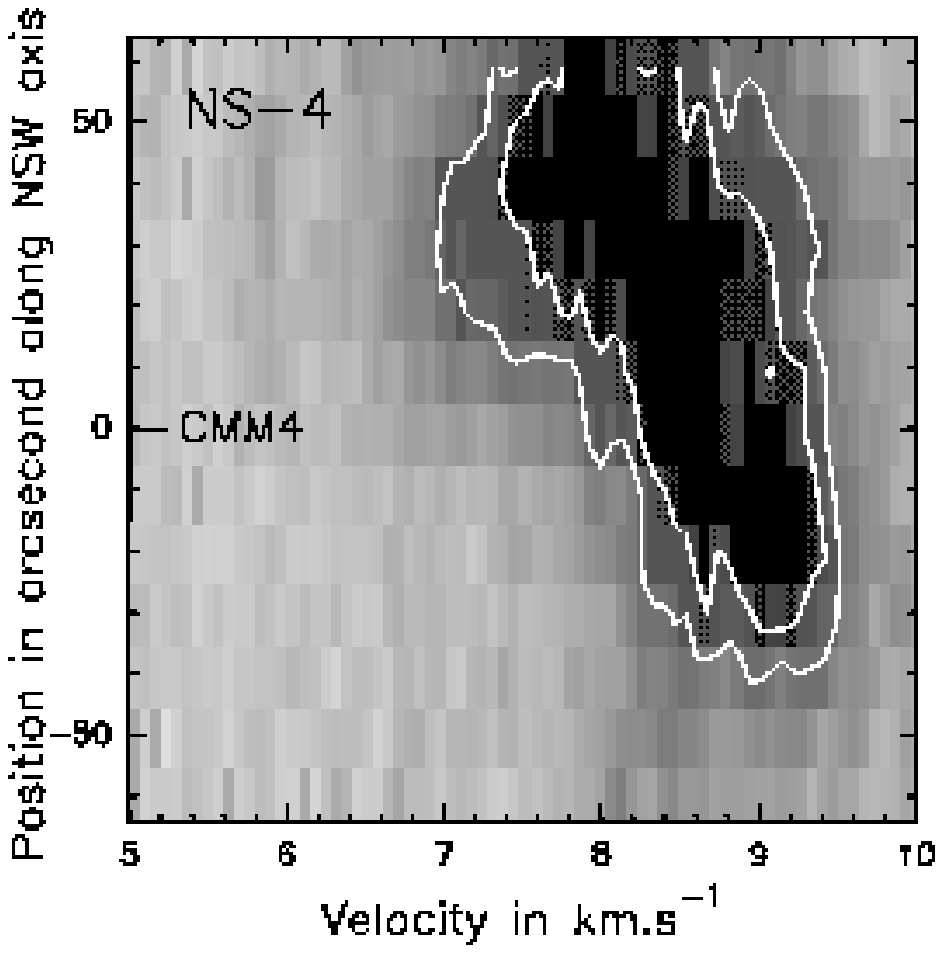,height=5.4cm,angle=0}}}
\vspace{-0.0cm}
\centerline{\hbox{
\hspace{-.5cm} 
\psfig{file=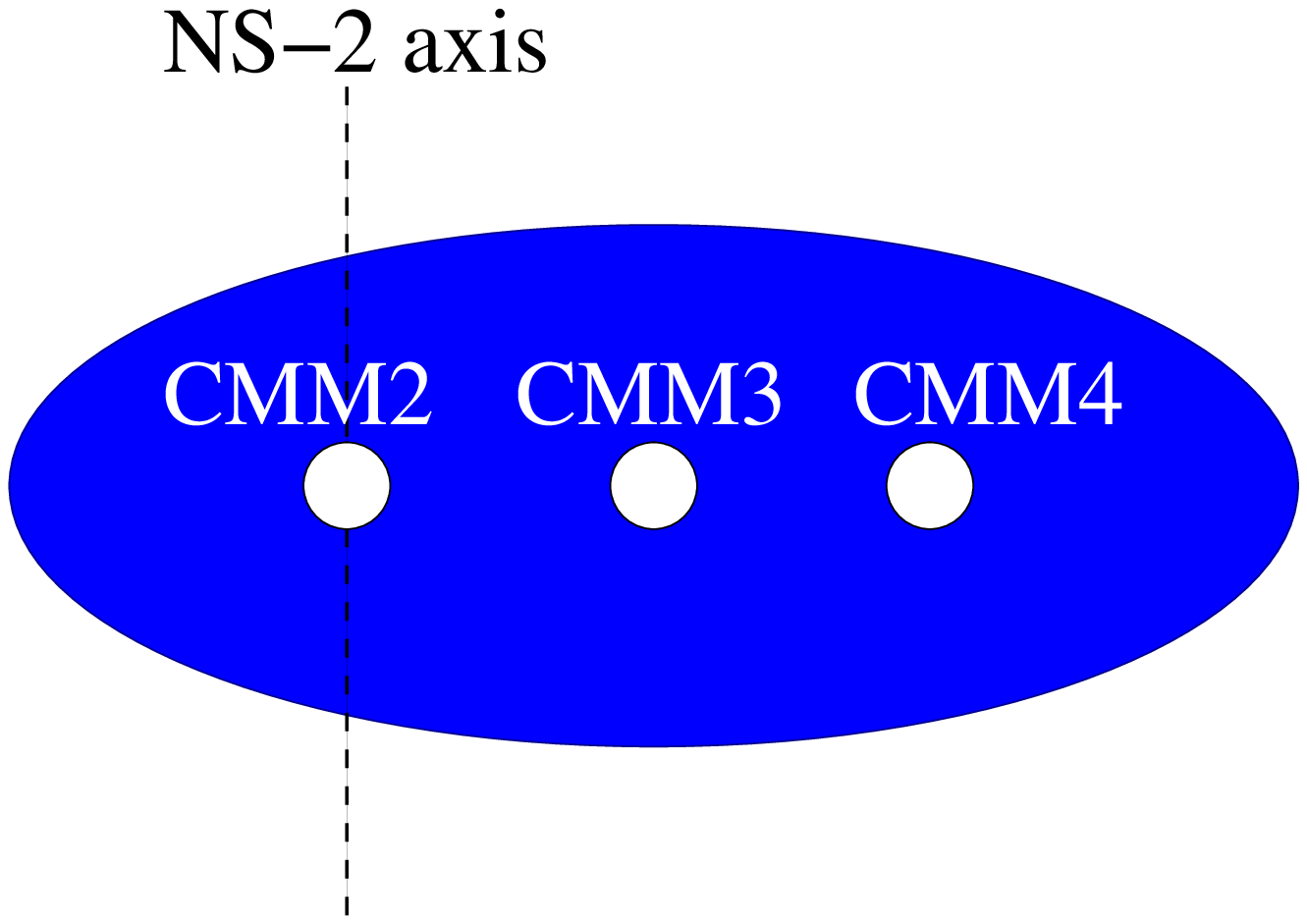,height=5.3cm,angle=270}
\hspace{-0.5cm}
\psfig{file=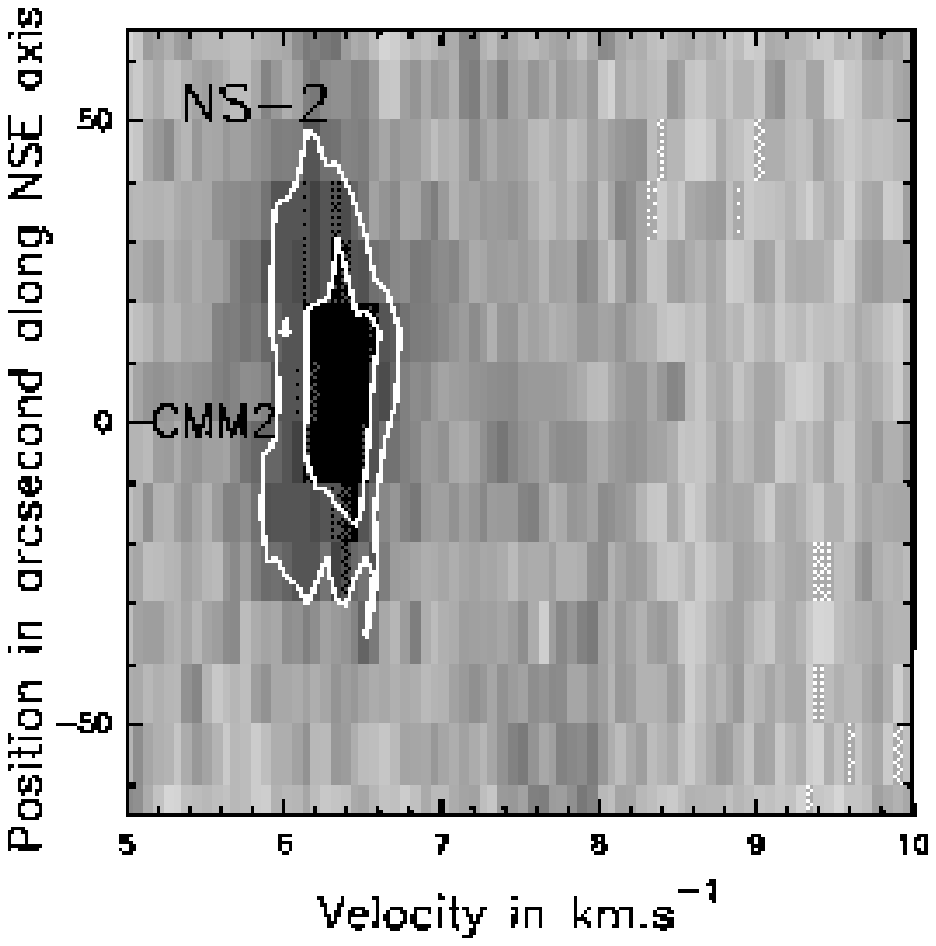,height=5.4cm,angle=0}
\hspace{0.2cm}
\psfig{file=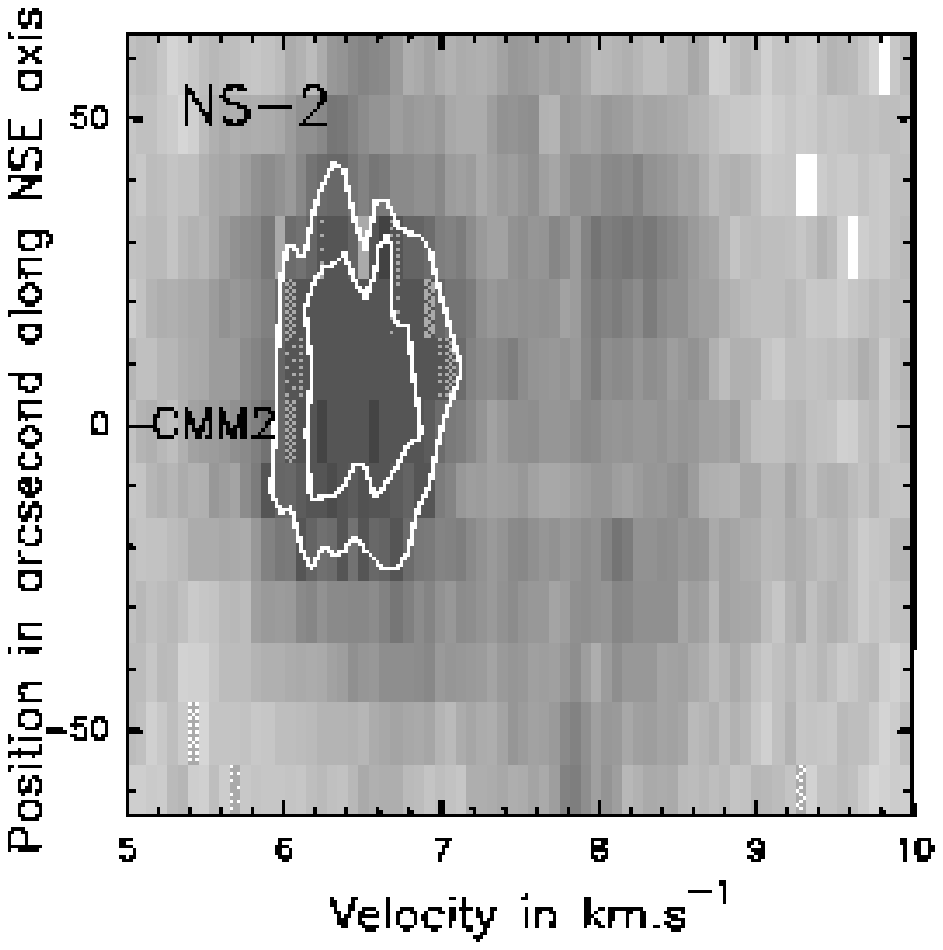,height=5.4cm,angle=0}}}
\caption{N$_2$H$^+$(101-012) and H$^{13}$CO$^{+}$(1-0) position-velocity diagrams observed 
in the central part of NGC~2264-C along four axes shown schematically on the left.
From top to bottom: the first row corresponds to PV diagrams taken along an East-West (EW) axis 
going through C-MM2, C-MM3, and C-MM4; the second row corresponds to a 
North-South axis going through C-MM3 (NS-3); 
the third row is for a North-South axis going through C-MM4 (NS-4); 
the fourth row is for a North-South axis going through C-MM2 (NS-2). Contour
levels are 1.1 and 1.8 K for the N$_2$H$^+$(101-012) diagrams,  
1.05 and 1.4 K for the H$^{13}$CO$^{+}$(1-0) diagrams.
\label{cmm2_posvel.ps}}
\end{figure*}

\subsection{A velocity discontinuity at the center of NGC~2264-C}
                                                                    
Our mapping in low-optical depth transitions (e.g. N$_{2}$H$^{+}$(1-0) and 
H$^{13}$CO$^{+}$(1-0)) revealed a remarkable velocity discontinuity in the
innermost part of NGC~2264-C, near the position of the continuum source C-MM3. 
This is illustrated in Fig.~\ref{cmm2_posvel.ps}, which shows 
N$_{2}$H$^{+}$(1-0) and H$^{13}$CO$^{+}$(1-0) position-velocity (PV) diagrams 
observed along four different axes.
In the PV diagrams taken along the East-West (EW) axis going through the
continuum sources C-MM2, C-MM3, C-MM4 (top two panels of 
Fig.~\ref{cmm2_posvel.ps}),
one can clearly see two distinct velocity components associated with 
C-MM2 and C-MM4 at $\sim$~6.5~km.s$^{-1}$ and $\sim$~8.5~km.s$^{-1}$, 
respectively, both extending over more than 50\arcsec  ~(i.e., $\sim$ 0.2 pc). 
These two velocity components overlap at the position of C-MM3, forming a 
sharp velocity discontinuity $\sim$ 2 km.s$^{-1}$ in amplitude. 
The other panels of Fig.~\ref{cmm2_posvel.ps} show PV diagrams taken 
along North-South axes at three different right-ascension positions, i.e., 
passing through C-MM2 (labelled NS-2 axis), C-MM3 (NS-3 axis), and C-MM4 
(NS-4 axis),
respectively. Along the NS-4 axis, only the higher velocity component at 
$\sim$ 8.5 km.s$^{-1}$ can be seen, while along the NS-2 axis only the lower 
velocity component at $\sim$ 6.5 km.s$^{-1}$ is visible. Both velocity 
components are visible in the PV diagram along the NS-3 axis.

We stress that the velocity feature seen in Fig.~\ref{cmm2_posvel.ps} cannot be 
explained by rotation. Indeed, both velocity components are strong at 
the center of the system (i.e. $\sim $ C-MM3) with little emission at
intermediate velocities, while the opposite trend would be expected in the 
case of rotation. Moreover, the rotational curve expected from differential rotation
is characterized by a continuous "S" shape (cf. Belloche et al. 2002) rather
than a sharp velocity discontinuity as observed here.

Finally, we note that no similar velocity discontinuity exists at the center 
of the NGC~2264-D clump.

\subsection{HCO$^+$(3-2) signatures of infall and outflow}

\begin{figure}[ht!]
\hspace{-0.7cm} 
\psfig{file=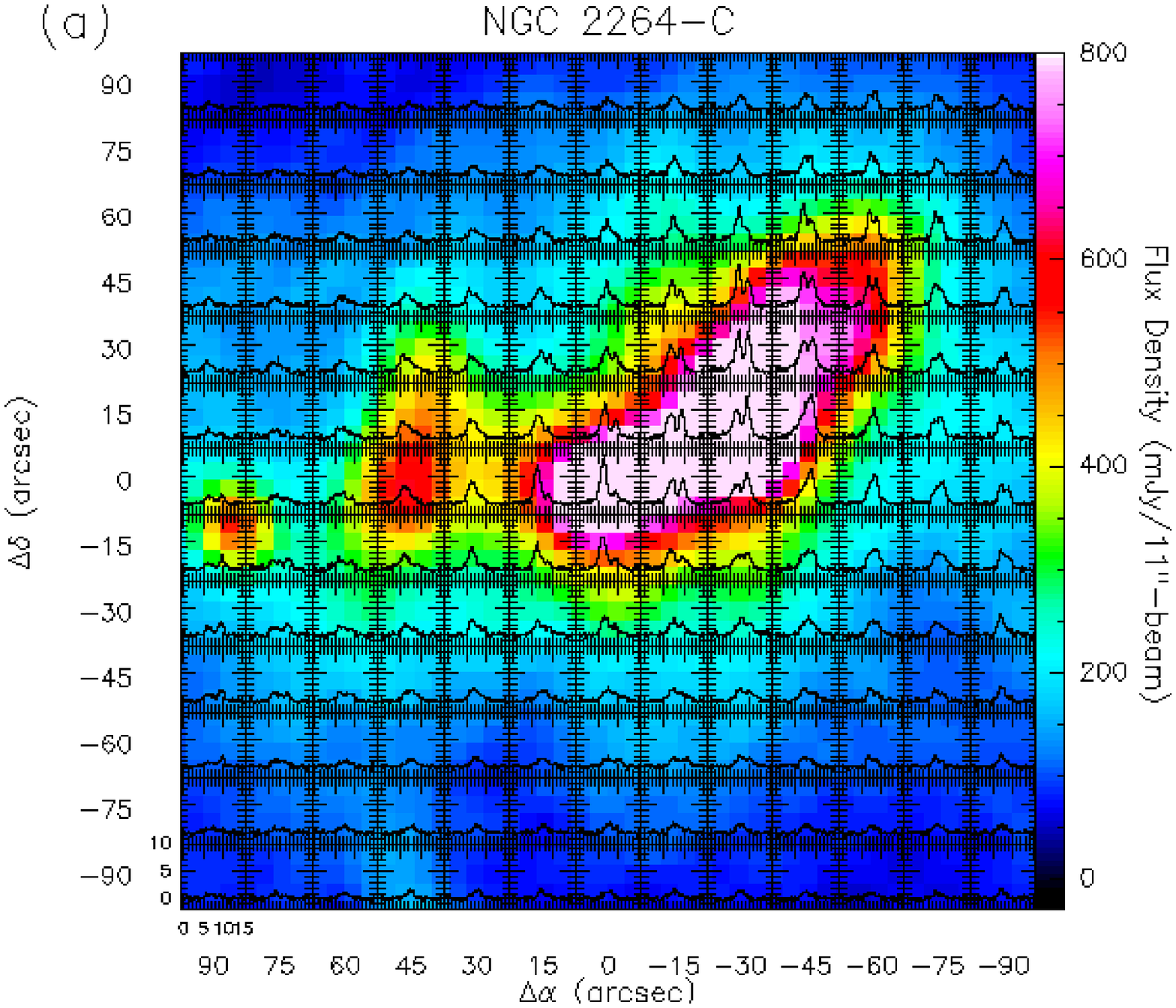,height=7.5cm,angle=0.}

\vspace{0.5cm}
\hspace{-0.7cm}
\psfig{file=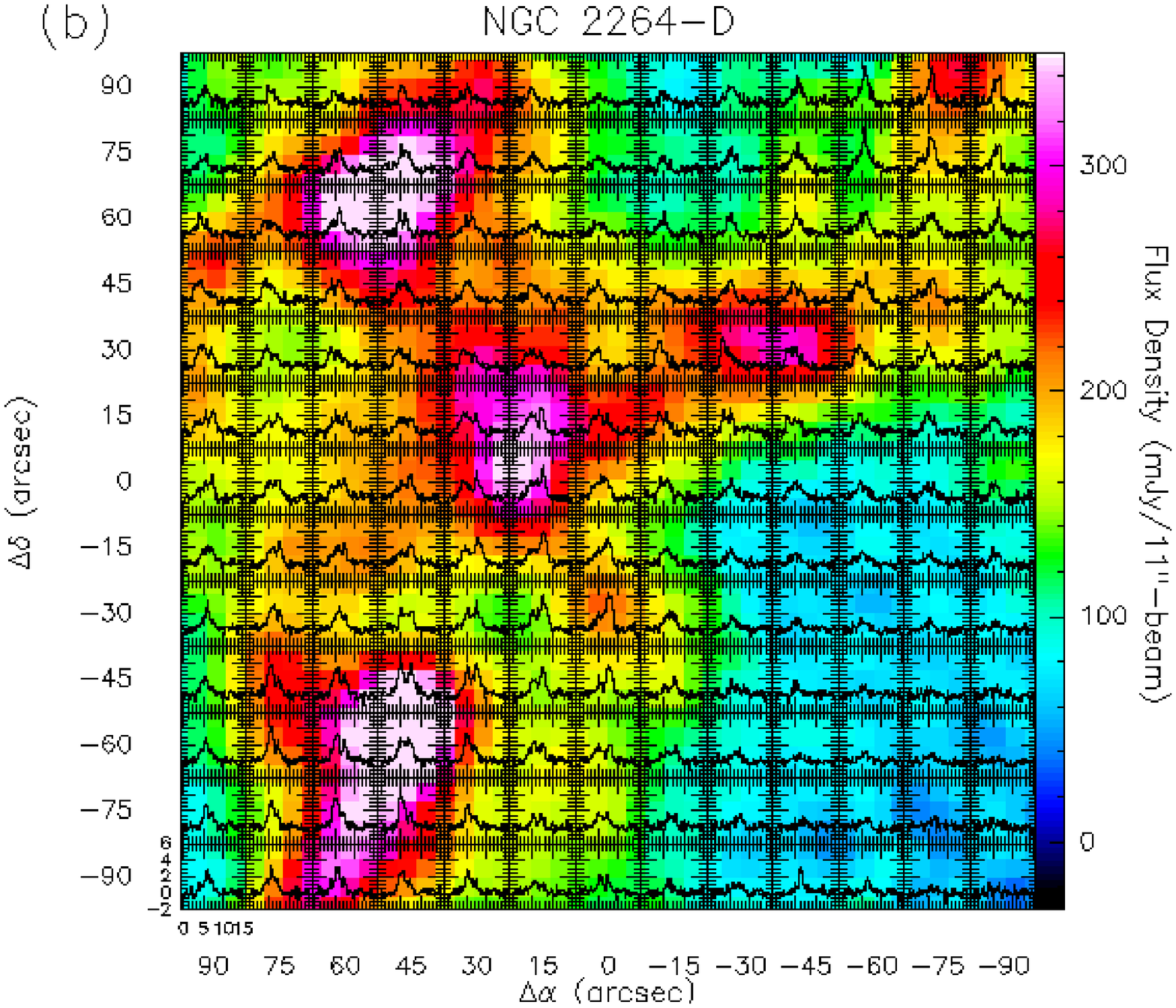,height=7.5cm,angle=0.}
\caption{\textbf{a)} Map of HCO$^{+}$(3-2) spectra observed toward NGC~2264-C, overlaid on the 1.2 mm dust continuum 
image (grey scale); 
the (0,0) position corresponds to C-MM3.
\textbf{b)} Same as (a) for NGC~2264-D;
the (0,0) position corresponds to D-MM3.
\label{spec-cont-ngc2264d.ps}}
\end{figure}

Our extensive HCO$^+$(3-2) maps of both NGC~2264 clumps are shown in 
Fig.~\ref{spec-cont-ngc2264d.ps}, in the form of spectra overlaid
on the 1.2 mm dust continuum images. 
The HCO$^+$(3-2) transition, which is optically thick and often self-absorbed 
in dense cores, is a good tracer of inward/outward motions (e.g. Evans 1999) 
when associated with an optically thin tracer such as H$^{13}$CO$^+$(3-2). 
A double-peaked HCO$^+$(3-2) spectrum with a blue peak stronger than the
red peak is usually taken to be a diagnostic of infall motions, providing
the corresponding optically thin H$^{13}$CO$^+$(3-2) spectrum
peaks in the dip of the HCO$^+$(3-2) line profile.
Conversely, an HCO$^+$(3-2) spectrum  
skewed to the red, with, e.g., a red peak stronger 
than the blue peak, diagnoses outflow motions. These diagnostics are valid
only for centrally-condensed sources (such as dense cores) as they assume 
that the line excitation temperature increases toward source center. 
In NGC~2264-C and NGC~2264-D, a variety of HCO$^+$(3-2) line shapes 
are observed (cf. Fig.~\ref{spec-cont-ngc2264d.ps}), ranging from 
clear `blue' infall profiles toward C-MM3 and D-MM1, to typical `red' outflow
profiles toward C-MM4 and D-MM3. Examples of unclear, mixed profiles are 
observed
toward C-MM2 or D-MM2. This variety of HCO$^+$(3-2) line shapes suggests that 
the velocity field within the NGC~2264 clumps is complex and possibly results from 
a combination of small-scale inflows/outflows directly associated with the 
MMSs and larger-scale systematic motions in the entire clumps. This will be 
discussed in \S~6 and \S~7 below.

\section{Nature of the embedded sources of NGC~2264}

We have classified the extracted MMSs as ``protostellar'' or ``prestellar'' 
based on the presence or absence of an outflow, jet or embedded YSO 
signature (see Table~\ref{resume_C} and Table~\ref{resume_D}). 
For this purpose, we use existing outflow/jet studies of the NGC~2264 region 
(Margulis et al. 1988; Schreyer et al. 1997, 2003; Wang et al. 2002), as well 
as the extensive 2MASS and MSX near-/mid-IR surveys. Our classification of the 
NGC~2264-D sources relies entirely on comparison with the 2MASS and MSX 
surveys as no systematic jet survey is available for clump D. 
By default, MMSs with no IR counterparts in NGC~2264-D are 
tentatively classified as 
``prestellar'', even though we cannot rule out that some may be Class~0 protostellar sources.

We find that up to $\sim$ 70$\%$ of the MMSs in clump C 
and at least 25$\%$ of the MMSs in clump D are protostellar in nature.
Although we generally lack good bolometric luminosity estimates for the protostellar MMSs 
and thus cannot prove that their submillimeter to
bolometric luminosity ratios match the definition of Class~0
protostars (Andr\'e et al. 1993), the fact they exhibit strong millimeter continuum emission suggests that they all are good
{\it candidate} Class~0 objects. We note that the Class~0 nature of one 
of the protostellar MMSs, D-MM1, has been reliably established 
on the basis of SCUBA imaging and HIRES processing of the $IRAS$ data   
(Wolf-Chase et al. 2003).

The apparent difference in protostellar fraction 
between clump C and clump D may be partly due to
the lack of outflow/jet survey in clump D. As an illustration, C-MM1 and C-MM2
in clump C are both classified as protostars because they show evidence 
of jets, while they are not associated with any near-/mid-IR sources.
The high fraction of protostars found in NGC~2264-C contrasts with the
significantly lower $\sim$ 40$\%$ fraction of protostellar sources observed 
by MAN98 in $\rho$-Oph.
It is also noteworthy that the protostellar sources of NGC~2264-C are all 
concentrated in the dense, inner ridge visible in Fig.~\ref{n2264D.ps}a, 
while the prestellar sources are all located in the outskirts of clump C.

When comparing our present observations of NGC~2264 with the study of MAN98
in $\rho$~Oph, one should keep in mind that there is a difference of a 
factor of $\sim 5$ in distance, hence in effective spatial resolution.
Based on the 1.2~mm continuum mosaic of MAN98, if $\rho$-Oph were at the 
same distance as NGC~2264 (800 pc), our source extraction method would only
have detected four sources (above the same detection threshold as in NGC~2264),
corresponding to the prominent dense cores Oph~A, Oph~B2, Oph~C, and Oph~F 
(cf. Loren, Wootten, \& Wilking 1990).
We conclude the MMSs identified here in NGC~2264 resemble more the DCO$^+$ 
'dense cores' of $\rho$~Oph than the compact starless condensations 
found by MAN98 (called `starless clumps' therein). This may partly
explain why we find a larger fraction of protostellar sources in NGC~2264: 
each dense core in $\rho$-Oph harbors at least one protostar and 
several starless condensations; observed from a distance of 800~pc, such a
core would likely be classified as `protostellar', even though the majority 
of its small-scale condensations actually are `prestellar'.

\begin{table}
\begin{minipage}[t]{\columnwidth}
\caption{Nature of the millimeter continuum sources identified in NGC~2264-C}    
\label{resume_C}      
\centering                          
\renewcommand{\footnoterule}{}
\begin{tabular}{c c c c c}        
\hline\hline 
Source   & MSX/2MASS\footnote{Number of 2MASS or MSX infrared sources lying within an 11\arcsec 
~beam centered on the MMS position} & H$_2$ jet\footnote{Detection of a shocked H$_2$ jet by Wang et al. (2002)} 
& CS flow\footnote{Detection of a CS outflow by Schreyer et al. (2003)} 
& Nature\footnote{'pro' stands for protostellar, 'pre' for prestellar} \\
\hline
C-MM1  &   0	  &  Y       &       & pro\\
C-MM2  &   0	  &  Y       &       & pro\\ 
C-MM3  &   3	  &  Y       &       & pro\\
C-MM4  &   1	  &  ?       & N     & pro\\
C-MM5  &   2	  &  ?       & N     & pro\\
C-MM6  &   0	  &  N       &       & pre\\
C-MM7  &   0	  &  N       &       & pre\\
C-MM8  &   0	  &  N       &       & pre\\
C-MM9  &   0	  &  N       &       & pre\\
C-MM10 &   1	  &  N       & Y     & pro\\
C-MM11 &   2	  &  N       &       & pro\\
C-MM12 &   2	  &  ?       & Y     & pro\\
\hline
\end{tabular}
\end{minipage}
\end{table}

\begin{table}
\begin{minipage}[t]{\columnwidth}
\caption{Nature of the millimeter continuum sources identified in NGC~2264-D}    
\label{resume_D}      
\centering                          
\renewcommand{\footnoterule}{}
\begin{tabular}{c c c }        
\hline\hline 
Source & MSX/2MASS\footnote{Number of 2MASS or MSX infrared sources lying within an 11\arcsec 
~beam centered on the MMS position} 
& Nature\footnote{'pro' stands for protostellar, 'pre' for prestellar} \\
\hline
D-MM1   &  1 	   & pro\\
D-MM2   &  0	   & pre\\
D-MM3   &  1	   & pro\\
D-MM4   &  0	   & pre\\
D-MM5   &  0	   & pre\\
D-MM6   &  0	   & pre\\
D-MM7   &  1	   & pro\\
D-MM8   &  0	   & pre\\
D-MM9   &  0	   & pre\\
D-MM10  &  0	   & pre\\
D-MM11  &  0	   & pre\\
D-MM12  &  0	   & pre\\
D-MM13  &  0	   & pre\\
D-MM14  &  1	   & pro\\
D-MM15  &  0	   & pre\\
\hline
\end{tabular}
\end{minipage}
\end{table}
On the other hand, millimeter continuum interferometer observations by Nakano 
et al.~(2003) and Schreyer et al.~(2003) show that the objects called 
C-MM4, C-MM5, and C-MM12 here remain unfragmented, single sources at a 
spatial resolution of $\sim$ 3000 AU.
This may suggest that the MMSs detected with the 30~m telescope in NGC~2264
are more centrally concentrated than the $\rho$-Oph dense cores (see \S ~8 
for further comparisons).\\ 

The bright infrared source IRS1 lies close to, but is clearly offset from, C-MM5
(Nakano et al. 2002; Schreyer et al. 2003). 
Based on its slightly rising spectral energy distribution between 
12~$\mu$m and 100~$\mu$m, Margulis et al. (1989) classified IRS1 as a Class~I 
object with L$_{bol}$=2300~L$_{\odot}$. The detections of VLA radio continuum 
emission (Schwartz et al. 1985), as well as H$_2$O and methanol masers, 
at the position of IRS1 are suggestive of a $\sim 9.5\, M_{\odot}$ 
YSO on the ZAMS (cf. Thompson et al. 1998). 
Although there is apparently no disk nor outflow directly associated with it
(Schreyer et al. 2003), IRS1 is thus clearly a relatively massive, embbeded 
young star.\\ 
The other infrared source detected by IRAS in the region discussed in this 
paper is IRS2, which lies in clump D, just outside the densest region 
(see Fig.~1 and Fig.~13). It is significantly less luminous than
IRS1 (L$_{bol}$=150~L$_{\odot}$ -- Margulis et al. 1989) and is not closely 
associated with any millimeter continuum source. 
It has also been classified as a relatively young (Class~I) object 
by Margulis et al. (1989).

\section{Radiative transfer modelling of NGC~2264-C}

Here, we compare our molecular line observations of NGC~2264-C 
with radiative transfer calculations performed with the same Monte-Carlo
radiative transfer code as used by Belloche et al. (2002). 
This code is divided into two parts. The first part calculates the non-LTE 
level populations based on a 1D Monte-Carlo method (Bernes 1978, 1979).
The second part, called MAPYSO (Blinder 1997), integrates the radiative 
transfer equation along each line of sight and convolves the resulting data 
cube with the beam of the telescope. MAPYSO can be used in either 1D 
or 2D geometry. The 2D mode was used for the present modelling of NGC~2264-C.
The inputs to the code are the density and abundance profiles, 
the kinetic temperature profile, the 2D velocity field, and the  
non-thermal velocity dispersion which, for simplicity, is assumed 
to be uniform over the simulated region.

The averaged radial density profile of NGC~2264-C can be inferred from 
the corresponding 1.2~mm dust continuum intensity profile I(r), where r is the spherical radius. 
The best power-law fit to the observed intensity profile is found to 
be I(r)~$\propto$~r$^{-0.4}$.  
Assuming isothermal dust emission in the Rayleigh-Jeans regime, this
translates into an average spherical density profile 
$\rho$(r)~$\propto$~r$^{-1.4}$
(cf. Motte \& Andr\'e 2001), which has been normalized so as to yield 
a total mass of 1650~M$_{\odot}$ within a radius of 0.4~pc, as observed (cf. Table~\ref{global} below).
Since the projected center of gravity of the NGC~2264-C clump as estimated from our 
dust continuum map is located at the position of C-MM3, we have centered our models
on C-MM3 rather than C-MM4 as adopted by Williams \& Garland (2002).\\
\begin{figure}[t!]
\vspace{-5cm}

\hspace{-8cm}
\psfig{file=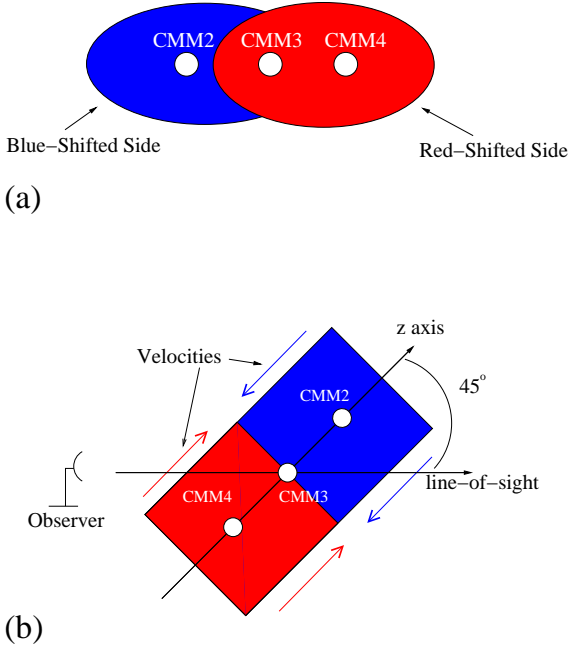,height=19cm,angle=270}

\vspace{-4cm}
\caption{
\textbf{a)} Schematic view of our proposed model of NGC~2264-C as seen by an 
observer in the 
plane of the sky.
\textbf{b)} Side view of the plane containing the z axis and the line of 
sight: 
a finite-sized, cylindrical clump collapses  
toward its center, coinciding with C-MM3. The long axis (i.e. z axis) of the 
cylinder makes an angle of 45 degrees 
with the line of sight.
\label{model2d.ps}}
\end{figure}
\begin{figure}[t!]
\hspace{1cm}
\psfig{file=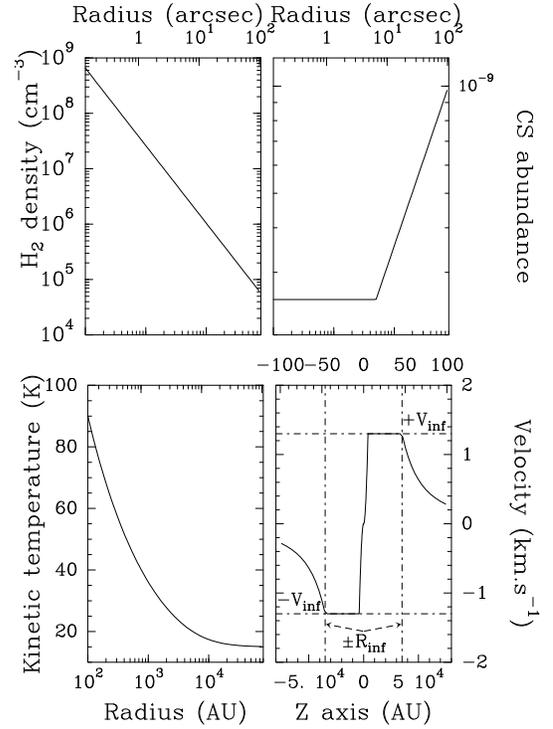,height=10cm,angle=0}
\caption{
 Profiles of the input parameters adopted in our 2D radiative 
transfer model of NGC~2264-C : Density scaling as $n \propto$ r$^{-1.4}$; 
depletion of the CS abundance by a factor of $\sim 5$ in the inner region
(the C$^{34}$S, H$^{13}$CO$^+$, and HCO$^+$ abundances are assumed to scale
 accordingly); 
gas kinetic temperature scaling as T$_k \propto$ r$^{-0.4}$ at small 
radii and fixed to T$_k$~=~15~K in the outer region; 
velocity profile along the z-axis of the cylinder (solid lines), with the locations of $\pm V_{inf}$ and 
$\pm R_{inf}$ marked by dash-dotted lines.
\label{model97.ps}}
\end{figure}

The gas kinetic temperature profile is taken to be 
$T_k(r) = ((T_0(\frac{r}{r_0})^{-0.4})^4 + T_1^4)^{0.25}$ with T$_0$=90 K, 
r$_0$=100 AU and T$_1$=15 K. The value of T$_0$ has been adjusted so as to match the intensity of the observed spectra.
This temperature profile is consistent with heating from a luminous central
source (L$_{bol} \sim 80$~L$_{\odot}$), radiating in an optically thin medium
and embedded in a
cold 15~K background at large radii.\\  
The CS and HCO$^+$ molecules are known to be depleted onto dust grains 
above a density $n_{H2} \sim$ 10$^{5}$ cm$^{-3}$ in starless cores (e.g. Tafalla et al. 2002). 
The spherical abundance profiles are thus fixed to standard values, i.e. [CS]/[H$_2$]=1$\times$10$^{-9}$ (i.e. Tafalla et al. 2002) and 
[HCO$^+$]/[H$_2$]=1.5$\times 10^{-9}$ (i.e. Bergin et al. 1997), 
in the low-density outer regions and are assumed to scale 
as $ n_{H_2}^{-0.4}$ with density (cf. Bacmann et al. 2002). 
We assume the decrease in abundance with density to stop  
where the kinetic temperature rises above 20~K in the inner region 
(see Fig.~\ref{model97.ps}). Still further in, where the gas becomes warmer than $\sim $~50~K, one may expects the abundances to increase as a result of grain mantle evaporation. However, T$_k$ reaches 50~K only at very small radii ($<$ 1000~AU corresponding to $<$ 1\arcsec) in our model, and this small inner region has no influence on spectra observed at $>$ 10\arcsec~angular resolution. 
We also adopt constant, standard 
values of 22 and 75 for the [CS]/[C$^{34}$S] and 
[HCO$^+$]/[H$^{13}$CO$^+$] isotopic ratios, respectively.\\ 
The non-thermal velocity dispersion is set to $\sigma_{NT}$=0.47 km.s$^{-1}$,
i.e., slightly less than the N$_2$H$^+$(1--0) linewidths measured in 
NGC~2264-C (cf. \S~ \ref{vel_global}), which accounts for the contribution of 
collapse motions to the linewidth. 
Finally, the infall velocity profile is adjusted so as to fit the observed spectra.

\begin{figure*}[ht!]
\begin{minipage}[c]{1.\textwidth}
\hspace{0.3cm}
\psfig{file=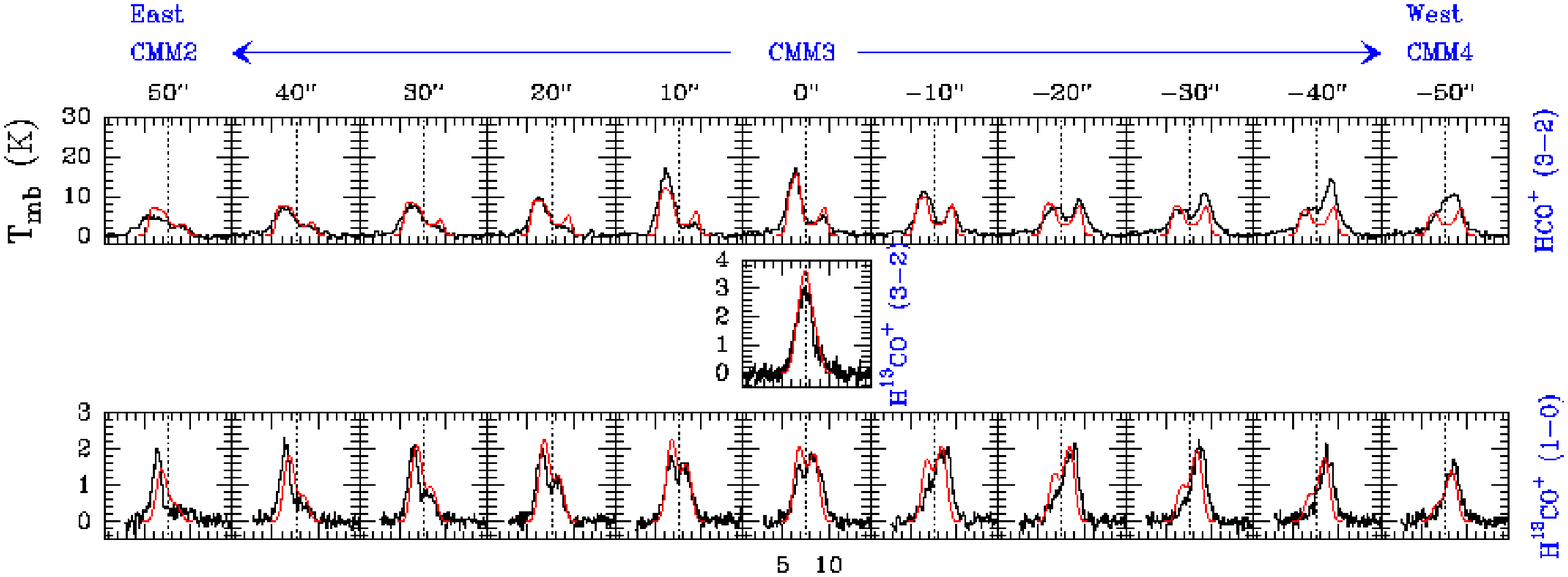,height=7.3cm,angle=0}
\end{minipage}
\begin{minipage}[c]{1.\textwidth}
\vspace{1cm}
\hspace{0cm}
\psfig{file=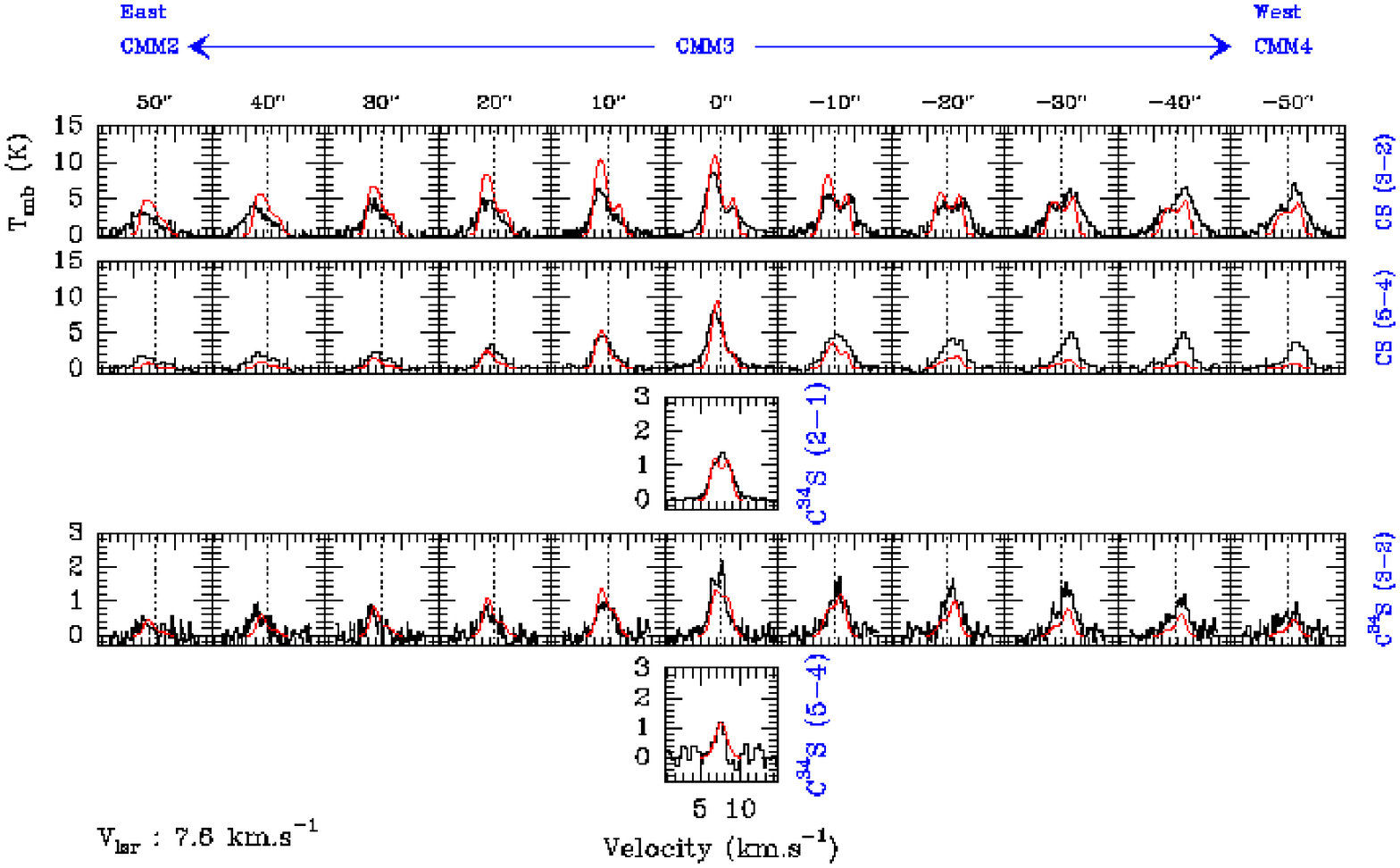,height=12cm,angle=0}
\end{minipage}
\hspace{1cm}
\vspace{-0cm}
\caption{HCO$^{+}$(3-2), H$^{13}$CO$^{+}$(3-2), H$^{13}$CO$^{+}$(1-0) (top)
and CS(3-2), CS(5-4), C$^{34}$S(2-1), C$^{34}$S(3-2), C$^{34}$S(5-4) (bottom) 
spectra observed along an East-West axis going through C-MM2, C-MM3, C-MM4 in 
NGC~2264-C. Synthetic spectra corresponding to the ``best-fit'' 2D radiative 
transfer model described in the text (see Fig.~\ref{model2d.ps}c for input 
parameters) are superimposed in light grey.
The central position corresponds to C-MM3.
\label{compare_ra0.ps}}
\end{figure*}
In order to account for the velocity discontinuity observed in the central part
of NGC~2264-C, we model the clump as a cylindrical filament centered 
on C-MM3 (see Fig.~\ref{model2d.ps}). 
The system is assumed to be collapsing toward its center of mass (i.e. C-MM3), 
so that the two sides of the filament (C-MM2 and C-MM4) are moving toward
each other. The long axis of the filament is taken to be inclined by 
45$^o$ to the line-of-sight (see Fig.~\ref{model2d.ps}b).
When probing the kinematics of such structure, one would indeed 
observe a sharp velocity discontinuity near the center of the filament. 
The velocity profile is taken to be only a function of z (position along 
the long axis of the cylinder), with v(z)=V$_{inf}$=1.3 km.s$^{-1}$ for z$<$0 
and v(z)=-V$_{inf}$ for z$>$0 in the inner $|z|<35000$~AU (or 45\arcsec) 
region, and sharp, symmetrical decreases of $|v(z)|$ with $|z|$ outside 
the central region (cf. Fig.~\ref{model97.ps}).  

Figure~\ref{compare_ra0.ps} compares the synthetic spectra calculated under  
these assumptions with the spectra observed along the EW axis in the CS(3-2),
CS(5-4), C$^{34}$S(2-1), C$^{34}$S(3-2), C$^{34}$S(5-4), HCO$^+$(3-2), 
H$^{13}$CO$^+$(3-2), and H$^{13}$CO$^+$(1-0)
transitions (where the (0\arcsec,0\arcsec) position corresponds to C-MM3). 
The agreement between the simulations and the observations is very encouraging 
given the simplicity of the model. 
The linewidths, peaks, and dips of the observed spectra are generally 
well reproduced. 
Furthermore, our simple cylindrical model manages to reproduce the 
observed reversal in line asymmetry, from blue-skewed spectra near C-MM3 and 
on the Eastern side of C-MM3 to red-skewed spectra on the Western side of C-MM3
(cf. Fig.~\ref{compare_ra0.ps}), without including any additional outflow 
velocity component near C-MM4. 
On the Western side of C-MM3, however, the intensity of the synthetic spectra 
is not strong 
enough, especially in the case of the optically thick HCO$^+$(3-2) and CS(5-4) 
transitions. This is because our 1D model underestimates both the actual 
density and the actual kinetic temperature, hence the 
excitation temperature, near the luminous IR source IRS1 and the millimeter
sources C-MM4 and C-MM5 (which are all offset from the center of gravity of 
the system at C-MM3).\\
We note that both the observed and synthetic H$^{13}$CO$^+$(1-0) spectra are 
asymmetric and even double-peaked near C-MM3. This is somewhat reminiscent 
of the asymmetry observed in HCO$^+$(3-2), although there is a fundamental
difference: While the HCO$^+$(3-2) spectra are optically thick and 
self-absorbed (with an optical depth $\tau \sim 45$ in the central velocity 
channel of the model), 
the H$^{13}$CO$^+$(1-0) transition is essentially optically thin 
(with $\tau \sim 0.6$ in the central velocity channel of the model), so that the 
asymmetric, double-peaked  H$^{13}$CO$^+$(1-0) profiles reflect the presence of 
two velocity components along the line-of-sight (see also the position-velocity 
diagrams shown in Fig.~\ref{cmm2_posvel.ps}) rather than self-absorption effects.

\begin{figure}[t!]
\vspace{-1cm}
\hspace{-1.8cm}
\psfig{file=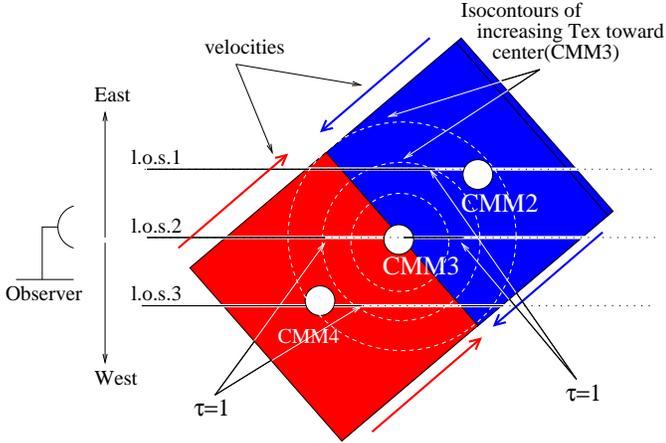,height=9.5cm,angle=270}
\vspace{-2cm}
\caption{Schematic view of the collapsing NGC~2264-C filament explaining the reversal of self-absorbed asymmetry 
observed in optically thick transitions (see \S~6). The positions of C-MM2, 
C-MM3, C-MM4 are marked.
The white dashed circles represent isocontours of increasing excitation temperature toward C-MM3.
Three lines of sight (l.o.s.) are shown and the point where the optical depth $\tau$~=~1 along each of them
indicated. 
\label{excitation.ps}}
\end{figure}

Figure~\ref{excitation.ps} provides a schematic explanation of the 
line reversal for optically thick spectra. Three lines of sight 
(l.o.s.~1, l.o.s.~2, and l.o.s.~3) are 
represented which intersect different isocontours of excitation temperature. 
Due to the elongated, cylindrical structure of the model, the blueshifted 
material probed along l.o.s.~3 has much lower excitation and/or column density
than the blueshifted material along l.o.s.~2 or l.o.s.~1. This explains why the 
intensity of the blue peak of the self-absorbed lines is strong on the 
Eastern side of C-MM3 and decreases rapidly on the Western side 
(see Fig.~\ref{compare_ra0.ps}).
By contrast, the two lines of sight l.o.s.~2 and l.o.s.~3 include similar 
amounts of excited, redshifted material, which qualitatively explains why the
 red peak remains approximately constant (in the model) on the Western side of 
C-MM3. 
Of course, protostellar outflows are likely to play an additional role in
shaping the line profiles observed in this region, especially around C-MM4 
and IRS~1 
(see the broad wings of the central CS(3-2) spectrum observed toward C-MM4 
in Fig.~\ref{compare_ra0.ps}).
Nevertheless, the above 
qualitative explanation of the asymmetry reversal and the reasonably good
model fit shown in Fig.~\ref{compare_ra0.ps} suggest that the velocity field
near C-MM3 is dominated by the global collapse of the entire NGC~2264-C 
elongated clump rather than by local inflowing/outflowing motions around
individual sources.\\ 

\begin{figure}[t]
\vspace{-2.0cm}
\hspace{-9cm}
\psfig{file=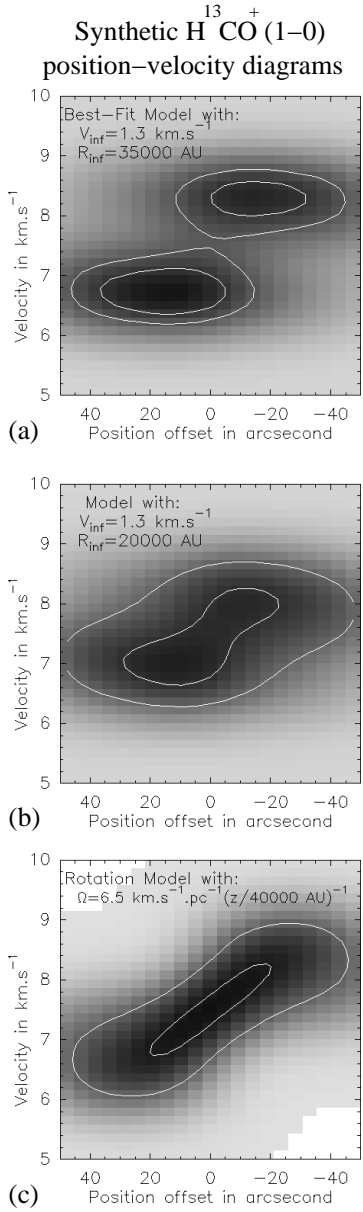,width=26.8cm,angle=270}
\vspace{-2.5cm}
\caption{Synthetic H$^{13}$CO$^+$(1-0) position-velocity diagrams calculated for three different 
radiative transfer models.
\textbf{a)} Diagram corresponding to the 'best-fit' model (see text for details). 
\textbf{b)} Diagram obtained by changing R$_{inf}$ to 20000~AU. \textbf{c)} Diagram obtained for 
a bulk (differential) rotation of the model cloud about an axis centered on C-MM3 and 
perpendicular to the line joining C-MM2, C-MM3, and C-MM4.
These diagrams have to be compared with the observed H$^{13}$CO$^+$(1-0) diagram shown at the 
top right of Fig.~\ref{cmm2_posvel.ps}.
\label{comp_model}}
\end{figure}

In order to estimate rough error bars on our 'best-fit' model 
shown in Fig.~\ref{compare_ra0.ps},
we have performed a set of radiative transfer calculations by varying some of the model parameters. 
We find acceptable fits for an inflow velocity V$_{inf}$=1.3~km.s$^{-1} \pm 0.2$~km.s$^{-1}$ and 
an inflow radius R$_{inf}$=35000~AU $\pm$ 10000~AU (see Fig.~\ref{model2d.ps}c for definitions of these 
two parameters). As an illustration, Fig.~\ref{comp_model} shows synthetic H$^{13}$CO$^+$(1-0)
position-velocity diagrams for three different models: (a) the ``best-fit'' model, (b) a model similar 
to the best-fit model but with R$_{inf}$=20000~AU  instead of R$_{inf}$=35000~AU , 
and (c) a model for which the velocity field corresponds to (differential) rotation about C-MM3 
as opposed to inflow toward C-MM3.
It can be seen that the agreement between the best-fit PV diagram (Fig.~\ref{comp_model}a) 
and the observed diagram (top right of Fig.~\ref{cmm2_posvel.ps}) is quite good, 
with two separate velocity components overlapping at the central position. 
By contrast, the model with a smaller inflow radius (Fig.~\ref{comp_model}b) does not match the 
main features of the observed PV diagram, even if an unresolved velocity discontinuity is apparent 
in the center. Likewise, the rotation model does not fit the observed PV diagram, producing 
a continuous velocity gradient as opposed to a velocity discontinuity. 
Based on the rotational models we have calculated, we can rule out rotation as the origin 
of the kinematic properties of NGC~2264-C. 
We also tried but failed to reproduce our observations with the kinematical model 
proposed by Williams \& Garland (2002), namely large-scale spherical infall 
onto an expanding central core coinciding with C-MM4. Our higher-resolution 
observations show that the large-scale collapse motions in NGC~2264-C 
converge toward a position much closer to C-MM3 than to C-MM4 and that these 
motions are not disrupted on small scales by the effect of protostellar 
outflows around C-MM4 as proposed by Williams \& Garland (2002).\\
Based on our best-fit model (cf Fig.~\ref{compare_ra0.ps}), the dynamical timescale 
of the inner part of NGC~2264-C is estimated to be 
t$_{dyn} \sim$ 1.7$\times$10$^5$ yr, which corresponds to the characteristic
time needed by C-MM2 and C-MM4 to collapse to the system central position 
(C-MM3). We can also calculate the mass inflow rate toward the center of 
the cylindrical filament 
$\dot{M}_{inf}= 2\times \pi~R^2_{fil} \times n_{mean} \times \mu \times m 
\times V_{inf}$, where R$_{fil}$ is the radius of the cylinder cross 
section, n$_{mean}$ is the mean number density in the cylinder, 
$\mu$ is the mean molecular weight, m is the mass of atomic hydrogen, and 
V$_{inf}$ is the inflow velocity used in our model fit. 
With R$_{fil}$~=~0.2~pc, 
n$_{mean}$~=~1$\times$10$^5$~cm$^{-3}$ (cf Table \ref{global}), $\mu$~=~2.33, 
V$_{inf}$~=~1.3~km.s$^{-1}$, we find 
$\dot{M}_{inf}~\sim$~3$\times$10$^{-3}$~M$_{\odot}$.yr$^{-1}$. This is 
an order of magnitude larger than the mass inflow rate found by 
Williams \& Garland (2002), which is not surprising since, in our picture,   
their spherical model was offset by $\sim 0.15$~pc from the true center 
of mass of NGC~2264-C.\\

\section{Radiative transfer modelling of D-MM1}

As the mean separation between nearest MMSs is larger in NGC~2264-D than in 
NGC~2264-C, our molecular line mapping observations of clump D should
be more sensitive to the dynamics of individual MMSs.
In this section, we present an attempt at modelling the strongest millimeter
source of NGC~2264-D, D-MM1, which was classified as a Class~0 protostar
by Wolf-Chase et al. (2003) and exhibits blue infall profiles in both 
HCO$^+$(3-2) and CS(3-2) (see, e.g., Fig.~\ref{spec-cont-ngc2264d.ps}).
\begin{figure}[t!]
\centerline{\hbox{
\begin{minipage}[c]{1.\textwidth}
\psfig{file=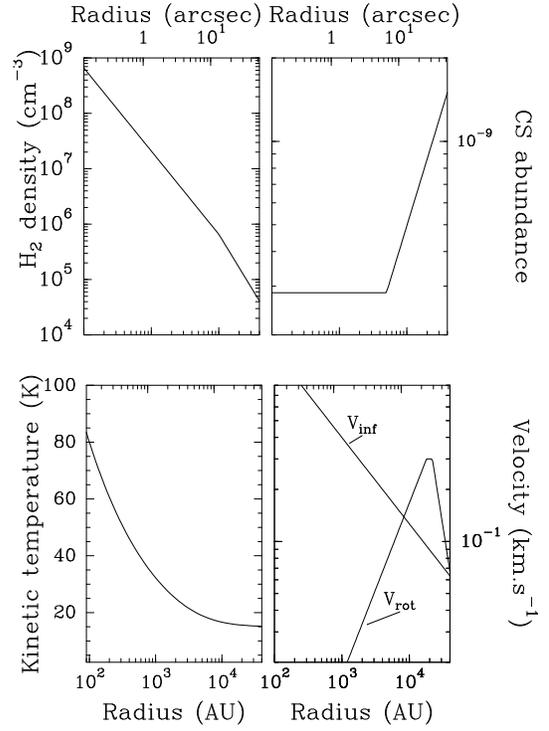,height=10cm,angle=00}
\end{minipage}}}
\caption{Radial profiles of the input parameters adopted in our 2D radiative 
transfer model of D-MM1: Density scaling as $n \propto$~r$^{-1.5}$ 
in the inner 10000~AU and $n \propto$~r$^{-2}$ beyond 10000~AU; depletion 
of the CS abundance by a factor of $\sim 6$ in the inner region (the profiles 
of the C$^{34}$S, 
HCO$^+$, and H$^{13}$CO$^+$ are assumed to have the same form); kinetic 
temperature scaling as T$_{k} \propto$ r$^{-0.4}$; 
infall velocity profile (marked V$_{inf}$) and rotational velocity
profile (marked V$_{rot}$ and depending only on the cylindrical radius from
the rotation axis)
\label{dmm1_model}}
\end{figure}

We have modelled D-MM1 as a spherical cloud core with both infall and rotational  
motions using the same radiative transfer code (MAPYSO) as for NGC~2264-C (\S ~6). 
Several input parameters of the code are well constrained. 
We assume the gas kinetic temperature to be well coupled to the 
dust temperature and adopt a dust temperature profile 
T$_{d} \propto$ r$^{-0.4}$ with T$_d \sim$ 80 K at r=100 AU 
(cf. Motte \& Andr\'e 2001 and references therein), 
consistent with the observed bolometric luminosity 
$L_{bol} \sim$ 100 L$_{\odot}$ (Wolf-Chase et al. 2003).
From our 1.2~mm continuum map, we derive a circularly-averaged radial flux 
density profile which is consistent with
 a radial density profile $\rho \propto$~r$^{-1.5}$ in the inner 
$1\times10^4$~AU radius region, and steeper outside with $\rho \propto$~r$^{-2}$. 
The model density profile has been normalized such that
 the mass of the cloud core is 10~M$_{\odot}$ within a radius of 
4500~AU as observed. 
The non-thermal velocity dispersion is set to $\sigma_{NT}$~=~0.51~km.s$^{-1}$
which is slightly less than the observed linewidth (cf Table \ref{table:4}). 
This is due to the collapse contribution to the linewidth.
  The N$_2$H$^+$(1-0) spectra we observe around D-MM1 indicate the
presence of a relatively strong velocity gradient 
$\sim$~3 km.s$^{-1}$.pc$^{-1}$ 
from south-west to north-east (P.A.$_{grad}~\sim$~+45$^o$). In agreement with 
our N$_2$H$^+$ data, the projection of the 
rotation axis on the plane of the sky has thus been fixed at 
P.A.$_{rot}~\sim -45^o$,
and an angular velocity $\Omega$=3 km.s$^{-1}$.pc$^{-1}$ has been adopted at 
$r = 2\times10^4$~AU from the center of D-MM1. 
 
\begin{figure}[ht!]
\vspace{-2cm}
\psfig{file=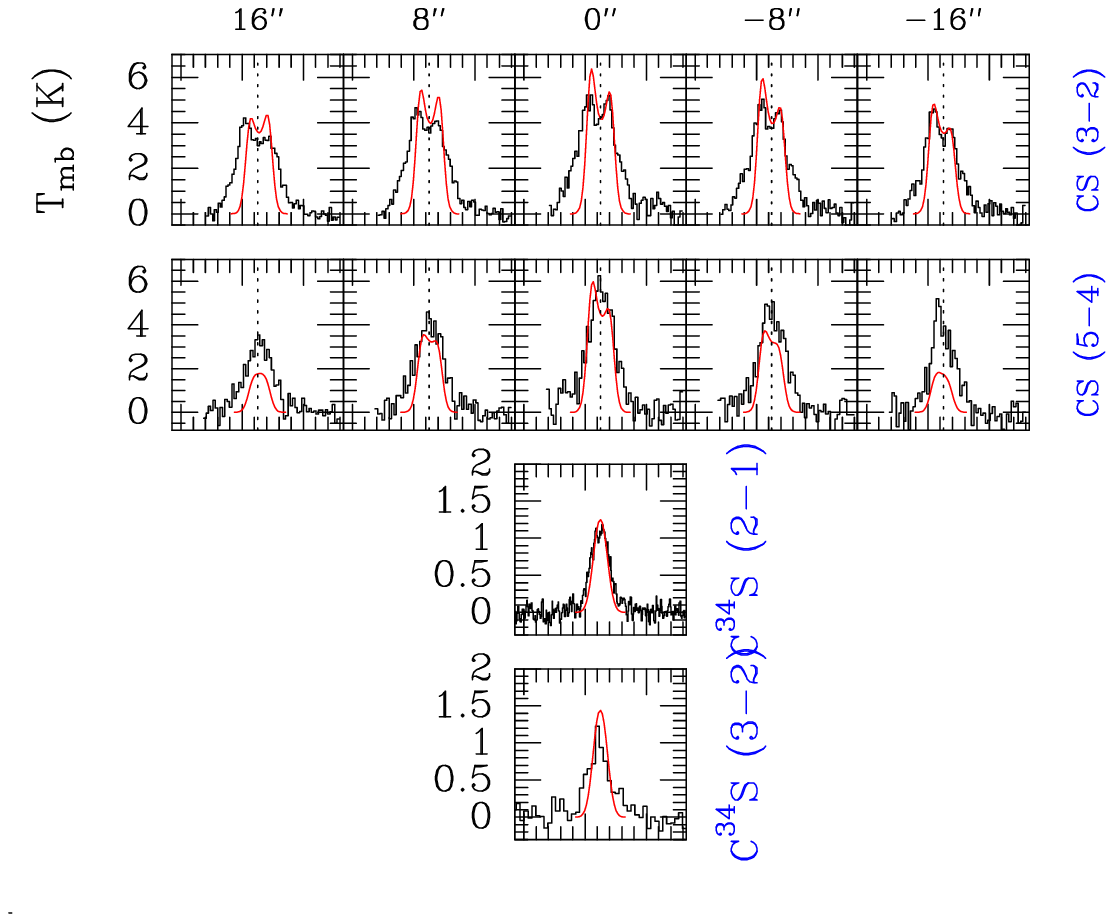,height=14.5cm,angle=270}
\vspace{-7.5cm}
\hspace{0.3cm}
\psfig{file=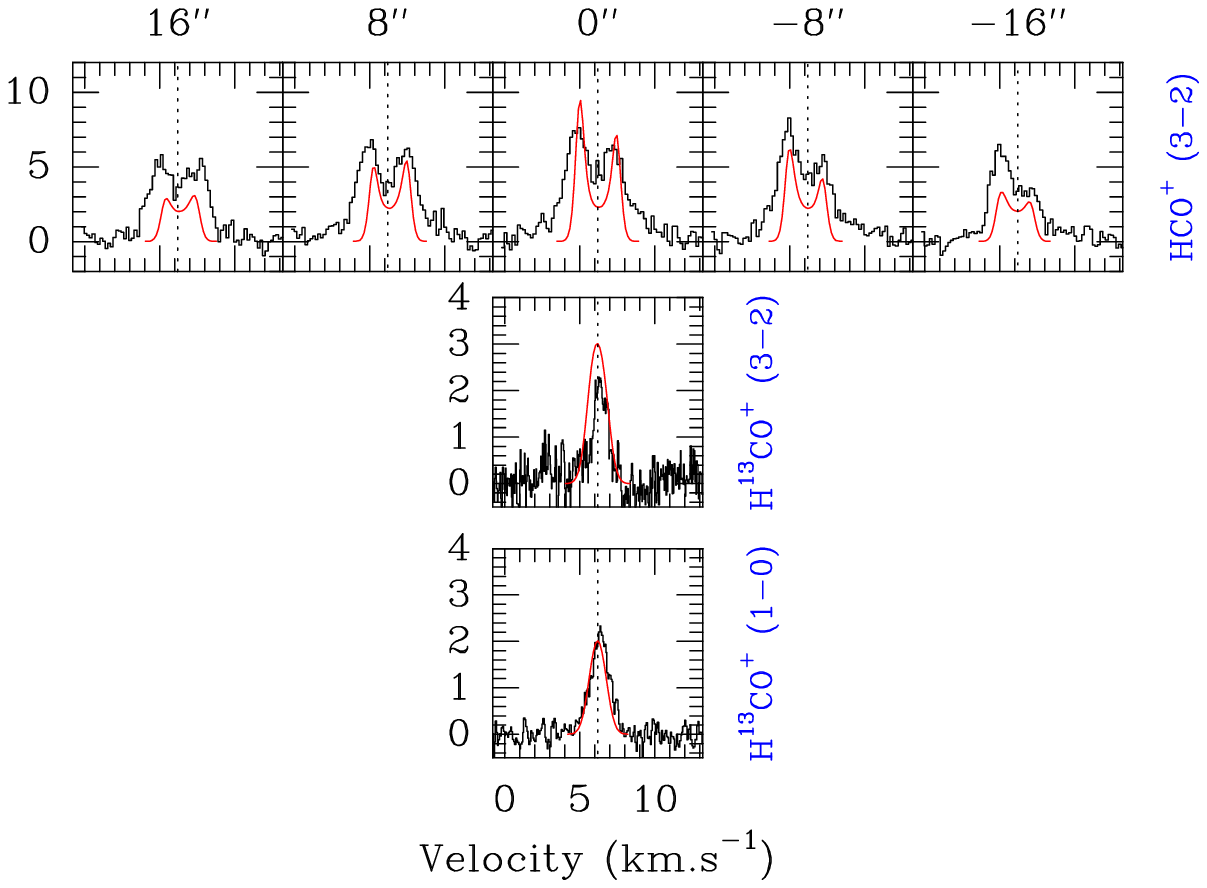,height=12.2cm,angle=270}
\vspace{-3cm}
\caption{ CS(3-2), CS(5-4), C$^{34}$S(2-1), C$^{34}$S(3-2) (top), and 
HCO$^{+}$(3-2), H$^{13}$CO$^{+}$(3-2), H$^{13}$CO$^{+}$(1-0) (bottom) spectra
observed along the direction of maximum velocity gradient (P.A.$_{grad}~\sim$~+45$^o$) 
through D-MM1. 
Synthetic spectra corresponding to the ``best-fit'' 2D radiative transfer model described 
in the text are superimposed in light grey. Positive offsets are located to
the North-East of D-MM1, while negative offsets are to the South-West.  
\label{dmm1_spec}}
\end{figure}

Taking the above constraints into account, the magnitudes of the 
infall and rotational velocities have been adjusted so as to match the
observations.
In the model shown in Fig.~\ref{dmm1_model} and Fig.~\ref{dmm1_spec}, 
the infall velocity is such as V$_{inf} \propto r^{-0.5}$ and fixed
to a value of 0.2~km.s$^{-1}$ at r~=~4000~AU. 
In order to reproduce the spatial sequence
of self-absorbed HCO$^+$ and CS spectra 
(clearly skewed to the blue at negative offsets
but almost symmetric at positive offsets -- see Fig.~\ref{dmm1_spec}), 
we find that strong differential rotation is required, with, e.g., 
solid-body rotation, V$_{rot} \propto r$, in the inner 
$2\times 10^4$~AU radius region and a sharp decline, V$_{rot} \propto r^{-2.5}$
in the outer region.
Such a rotational velocity profile is reminiscent of the velocity field found 
by Belloche et al. (2002) for the low-mass Class~0 object IRAM 04191 in Taurus.
With these parameters and assumptions (summarized in Fig.~\ref{dmm1_model}), 
a reasonably good fit is found (see Fig.~\ref{dmm1_spec}), although the model 
fails to reproduce the broad wings observed in HCO$^+$(3-2) and CS(3-2), 
probably due to outflowing material.  

Based on this model fit, we can estimate the mass infall rate onto D-MM1 
as follows:
$\dot{M}_{DMM1}$~=~4$\pi$R$^2~\times~n_{H_2}~\times~\mu~\times~m~\times$~
V$_{inf}$. 
At a radius of 10000~AU,  V$_{inf}$~$\simeq$~0.1 km.s$^{-1}$ and 
n$_{H_2} \sim$~6$\times$10$^5$~cm$^{-3}$, which yields  $\dot{M}_{DMM1}$~=~1.1$\times$10$^{-4}$~M$_{\odot}$.yr$^{-1}$.
This is almost two orders of magnitude larger than the typical infall 
rate measured toward low-mass protostars in Taurus 
(Ohashi 1999, Belloche et al. 2002), 
and comparable to the large infall rate estimated 
for some Class~0 protostars in protoclusters such as NGC~1333-IRAS4A 
(cf. Di Francesco et al. 2001, Andr\'e et al. 2004).
The implications of such a high infall rate are discussed in the next 
section.

\section{Discussion and Conclusions}

\subsection{Comparison of three protoclusters: NGC~2264-C, NGC~2264-D, and 
$\rho$-Ophiuchi}

The overall dynamical state and stability of a clump may influence the 
formation process of individual stars in its interior.
We have shown in \S ~6 that the NGC~2264-C clump is in a state of 
global collapse and 
evolves on the dynamical timescale t$_{dyn} \sim$ 1.7$\times$10$^5$ yr. 
This implies that NGC~2264-C is {\it not} in approximate hydrostatic 
equilibrium. 
Independently, we can compare the 3D velocity dispersion of the MMSs, $\sigma _{3D}$, 
to the 3D velocity 
dispersion expected in virial equilibrium, $\sigma _{vir\_ 3D}$.
These parameters are given in Table \ref{global} for both NGC~2264-C and 
NGC~2264-D, as well as $\rho$-Ophiuchi, the nearest ($d \sim 150$~pc)
example of a cluster-forming clump.
Table \ref{global} also lists the free-fall time of each clump estimated
from the observed mean volume density. 
In all three cases, the observed velocity dispersion of the MMSs 
appears to be insufficient to balance gravity. 
Likewise, the estimated virial mass M$_{vir}$ is significantly smaller 
than the total gas mass derived from the 1.2~mm continuum map for both clumps
(cf. Table \ref{global}), corresponding to a virial parameter 
$\alpha_{vir} = M_{vir}/M_{1.2} \simeq 0.2$, which suggests that
NGC~2264-C and NGC~2264-D are both out of virial equilibrium 
and highly unstable to collapse. If the clumps are gravitationally 
unstable, we expect the free-fall time to provide a reasonably good estimate
of the characteristic evolutionary timescale. 
Our independent estimate of t$_{dyn}$ for NGC~2264-C (see above) supports this 
view since t$_{dyn}$ is only slightly larger than t$_{ff}$. 
We conclude that the central part of NGC~2264-C is close to being in free-fall 
collapse. 
NGC~2264-D and $\rho$-Ophiuchi may also be collapsing.
Using HCO$^+$(3--2), 
Williams \& Garland (2002) found some evidence for large-scale collapse 
in NGC~2264-D at a characteristic velocity v$_{coll.}$~=~0.3~km.s$^{-1}$, 
although more observations of optically-thick tracers on large scales would be 
required to confirm their result. 
Likewise, tentative signatures of large-scale infall have been observed in 
$^{12}$CO and $^{13}$CO toward the $\rho$~Oph clump  
(e.g. Encrenaz et al. 1975, Wilking \& Lada 1983).\\ 

\begin{table*}
\begin{minipage}[t]{\textwidth}
\caption{Global properties of NGC~2264-C and NGC~2264-D compared to those of $\rho$-Oph} 
\label{global}      
\centering                          
\renewcommand{\footnoterule}{}
\begin{tabular}{c c c c c c c c c c}        
\hline\hline                 
Cl. Name  & Cl.Diam.\footnote{Mean clump diameter measured at the 70 mJy/11\arcsec-beam contour in the 1.2 mm continuum map (cf. Fig.~\ref{n2264D.ps}a and Fig.~\ref{n2264D.ps}b)}
 & Cl.Mass\footnote{Total gas mass of the clump estimated from the 1.2~mm continuum image prior to spatial filtering} 
& M$_{vir}$\footnote{Virial mass calculated in the same way as in Table \ref{table:4} from the mean N$_2$H$^+$(1-0) 
line-of-sight velocity dispersion observed in the clump (cf. col.[3] of Table~\ref{resume_vel})}
& Cl.Col.Dens.\footnote{Mean column 
density averaged over each clump} & Cl.Vol.Dens.\footnote{Mean density
 of each clump assuming a spherical volume}& t$_{ff}$\footnote{Mean free-fall 
time of the clump} & $\sigma _{3D}$\footnote{Observed 3D velocity dispersion of the MMSs in each clump} & 
$\sigma _{vir\_ 3D}$\footnote{Expected 3D velocity dispersion of the MMSs if they were virialized 
in the gravitational potential of the clump: $\sigma _{vir\_ 3D} = \sqrt{\frac{G~M_{1.2}}{R}}$ }\\
& (pc) &  (M$_{\odot}$) & (M$_{\odot}$) &(10$^{22}$cm$^{-2}$) & (10$^{4}$cm$^{-3}$) & 
(10$^5$yr)  & (km.s$^{-1}$) & (km.s$^{-1}$)\\
\hline 
NGC2264-C & 0.8 & 1650 & 340  &16 & 10.7 &  1.0 & 1.3  & 3.7   
\\
NGC2264-D & 0.9 & 1310 & 310 &11 & 6.0 &  1.4 & 1.4  & 3.0  
\\
\hline
$\rho$-Oph\footnote{Based on the C$^{18}$O(1-0) results of Wilking \& Lada (1983) for the diameter and the mass, 
and on Belloche et al. (2001) for the velocity dispersion} & 1.1 & 550 & -- & 3 & 1.4 &  2.9  & 0.6 & 2.1 
\\
\hline
\end{tabular}
\end{minipage}
\end{table*}

\begin{figure}
\hspace{-0.cm}
\psfig{file=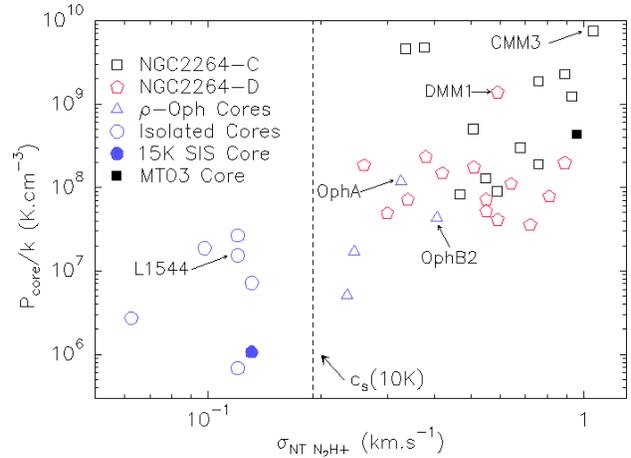,height=6.0cm,angle=0}
\caption{Plot of mean pressure versus non-thermal velocity dispersion for the millimeter
continuum sources identified in NGC~2264-C (square symbols) and NGC~2264-D (pentagonal symbols), 
the $\rho$~Oph DCO$^+$ cores Oph~A, OphB~2, Oph~C, Oph~F (open triangles -- from MAN98 and 
Belloche et al. 2001), and the isolated prestellar cores L1512, L1544, L183, L1696A, L63, L1155C
(open circles -- from Ward-Thompson et al. 1999 and Caselli et al. 2002). 
The location of a singular isothermal sphere (SIS) model with an effective temperature of 
15~K and an actual gas temperature of 10~K is shown as a filled circle for comparison. The filled square corresponds to the turbulent core model of 
McKee \& Tan (2003 -- MT03) for a 
clump surface density $\Sigma_{cl}=1$~g.cm$^{-2}$
and a final stellar mass $m_{\star f} =10$~M$_{\odot}$ (see also \S~8.2).
The vertical dashed line separates cores dominated by thermal motions (on the left) 
from cores dominated by supersonic non-thermal motions (on the right), assuming a typical 
gas temperature of 10~K.
\label{press_sigma}}
\end{figure}

We now compare the MMSs of NGC~2264 with the DCO$^+$ cores of 
$\rho$~Oph (Loren et al. 1990) and typical isolated prestellar cores 
(Ward-Thompson et al. 1999, Caselli et al. 2002). 
Figure~\ref{press_sigma} shows a plot of mean pressure, 
$\bar{P}_{core}/k$, versus 
non-thermal velocity dispersion, $\sigma_{NT}$, for these various types of 
cores as well as a typical singular isothermal sphere (SIS) model 
(Shu et al. 1987) and a typical turbulent core model (McKee \& Tan 2003 -- 
hereafter MT03). 
The mean pressure in each core has been calculated from its 
mean column density derived from millimeter continuum observations,  
using Eqs.~(A5-A6) of MT03 and assuming $\alpha_{vir}=1$ 
(which is typical of the  NGC~2264 MMSs -- see Table~\ref{table:4}). 
The non-thermal component of the 1D velocity dispersion in each core
has been estimated from the observed N$_2$H$^+$(101-012) linewidth after 
subtracting the thermal broadening expected for T$_k$=10~K.\\ 
It can be seen in Fig.~\ref{press_sigma} that $\bar{P}_{core}/k$ increases 
by three orders of magnitude and $\sigma_{NT}$ increases by one order
of magnitude from the thermally-dominated, isolated prestellar cores 
(cf. Myers 1998) to the most extreme MMSs of NGC~2264 (this paper)
dominated by supersonic non-thermal motions.
It can also be noted that the hydrostatic SIS model provides a reasonably good 
representation of the locations of isolated prestellar cores in the 
$\bar{P}_{core}/k$~vs.~$\sigma_{NT}$ diagram, while the turbulent core model 
of MT03 (see \S ~8.2 below) provides a much better description of the 
NGC~2264 sources. 
Interestingly, the low-pressure, thermally-dominated cores on the 
left-hand side of Fig.~\ref{press_sigma} are typically forming one (or two) 
stars, while the high-pressure, ``turbulent'' cores on the right-hand side 
are associated with protoclusters. A similar distinction between the 
``isolated'' and ``clustered'' modes of star formation was noted earlier 
by Myers (1998) and Jijina et al. (1999).\\
In the high-pressure, high-$\sigma_{NT}$ part of Fig.~\ref{press_sigma}, 
it can be seen that the $\rho$-Oph cores Oph~A and Oph~B2 are located in 
the regime of the MMSs of NGC~2264-D. The sources of NGC~2264-C have higher 
mean pressure and larger non-thermal velocity dispersions by a factor 
$\sim 10$ and $\sim 2$, respectively, on average than OphA and OphB2.
The most extreme source of NGC~2264-C (C-MM3) has a mean pressure 
$\bar{P}_{core}/k \sim 10^{10}\, \rm {K\, cm}^{-3}$ reminiscent of the 
pressure found in high-mass star-forming clumps (cf. MT03). 

We conclude that the MMSs of NGC~2264, especially those of clump C, 
represent somewhat more pressurized and more turbulent conditions for star 
formation than do the $\rho$~Oph cores. 
The more extreme conditions found in NGC~2264-C are actually  
reminiscent of those suggested by MT03 for the precursors of massive stars.

\subsection{Comparison with two scenarios of clustered star formation}

\begin{figure}
\vspace{-0.0cm}
\hspace{-0.5cm}
\psfig{file=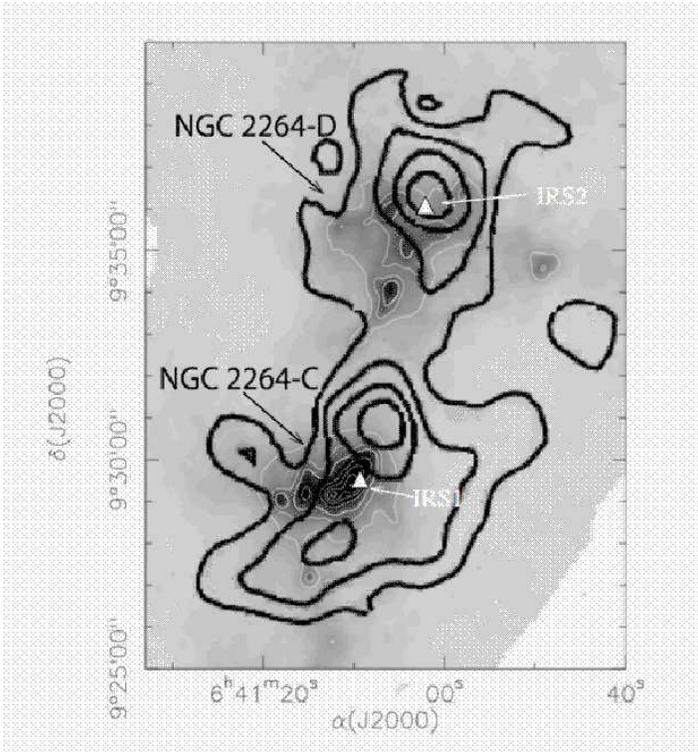,height=10cm,angle=0}
\vspace{-0cm}
\caption{Contour map of the surface density of near-IR sources in the young star cluster 
associated with NGC~2264 (from Lada \& Lada 2003), overlaid on our 1.2~mm continuum 
mosaic of the region (greyscale and white contours). 
The white filled triangles mark the positions of IRS1 and IRS2.
Note that the maxima of the near-IR source surface density distribution are located 
{\it outside} the dense gas/dust clumps NGC~2264-C and NGC~2264-D, while the projected
position of the luminous near-IR source IRS1 lies {\it within} the 1.2~mm contours 
of clump~C   
\label{lada.ps}}
\end{figure}

The young near-infrared star cluster
associated with NGC~2264 (Lada et al. 1993) is known to be 
highly structured and exhibits a surface density
distribution with multiple peaks (see Fig.~4 of Lada \& Lada 2003).
Two of these peaks lie close to, but are clearly offset from, the
cluster-forming clumps NGC~2264-C and NGC~2264-D studied in the present 
paper (see Fig.~\ref{lada.ps}).
This is suggestive of at least two different episodes of (clustered) 
star formation in the NGC~2264 region.
The detection of a few 2MASS and MSX infrared sources toward both NGC~2264-C 
and NGC~2264-D shows that both clumps have already formed YSOs. 
Furthermore, the presence of the massive (B2-type) Class~I infrared YSO IRS1 
within the NGC~2264-C clump sets interesting constraints on the process of 
intermediate- to high-mass star formation.
 
According to the scenario proposed by Bonnell et al. (1998, 2002), 
the formation of massive stars occurs through competitive accretion 
and stellar mergers
in the dense inner core of contracting protoclusters when their central 
stellar density exceeds $\sim$ 10$^8$~stars/pc$^{3}$. 
The gravitational potential well of such cluster cores is dominated by the stellar
rather than the gas component.
We find that it is difficult to account for the formation of IRS1 in the 
context of this 
scenario. Indeed, IRS1 appears to be located in the outer parts of the 
NGC~2264-C gas clump (e.g. Thompson et al. 1998, Nakano et al. 2002), 
rather near the peak of the near-IR source density distribution 
(cf. Fig.~\ref{lada.ps}). 
Figure~1 from Lada et al. (1993) indicates a stellar surface density $\sim 80$~pc$^{-2}$,
which allows us to estimate the stellar contribution to the gravitational potential 
within the volume of NGC~2264-C. 
Assuming the Kroupa (2001) IMF, for which the mean stellar mass is $\sim 0.4$~M$_{\odot}$,
the above number surface density translates into a stellar mass surface density 
$\sim$~30~M$_{\odot}$.pc$^{-2}$, i.e., less than 2$\%$ of the total gas mass surface density. 
Even allowing for a possible fraction of highly extinguished stars in the protocluster 
that would not be included in the above estimate, we conclude that the gravitational potential 
of NGC~2264-C is dominated by the gas rather than by the stars.
Thus, IRS1, the most massive infrared YSO in this region
is unlikely to have formed by collision and coalescence of lower-mass {\it stars}
as described by Bonnell et al. 
On the other hand, the global, near free-fall collapse of NGC~2264-C may lead 
to strong dynamical interactions between MMSs and even {\it protostellar} 
mergers near the center of the gas-dominated clump (see below).

In the alternative scenario proposed by McKee \& Tan (2002, 2003 --  
e.g. MT03), high-mass stars form from massive, 
turbulent cores embedded in a high-pressure, environment/clump. 
Both the cores and the ambient medium are assumed to be centrally-concentrated 
and initially virialized, i.e., close to hydrostatic equilibrium. 
Although clumps C and D are indeed massive and turbulent, our observations 
reveal that they are not in hydrostatic equilibrium but rather in a state of 
global dynamical collapse, at least in the case of NGC~2264-C. 
The primary characteristic of the MT03 model, however, is that the prestellar 
precursors of massive protostars are high-density, turbulent cores, which 
leads to large, accelerating accretion rates during the protostellar phase.
Large ($\simgt$~10$^{-4}$~M$_{\odot}$.yr$^{-1}$) accretion rates are  
indeed necessary to overcome radiation pressure and allow high-mass 
($M_\star \simgt 8\, M_\odot $) stars to form by accretion despite the 
high luminosity of the central stars (e.g. Wolfire \& Cassinelli 1987).
Large-scale collapse can help to produce dynamically pressurized, high-density
cores in the inner part of massive cluster-forming clumps. 
Even though one of the premises of the MT03 model is not verified, a detailed 
comparison of the model with the observed properties of NGC~2264-C and its 
central source, C-MM3, is therefore still warranted.
 
The mean column density of the NGC~2264-C clump is 
$N_{H2} \sim 1.6 \times 10^{23}\, \rm{cm}^{-2}$, corresponding to a mean 
surface density $\Sigma_{cl} \sim$~0.6~g.cm$^{-2}$, typical of that 
observed in most regions of massive star formation and consistent 
with the assumptions of the turbulent core model of MT03.
With a mass of 40~M$_{\odot}$ and a line velocity dispersion 
$\sigma _{line}$~=~1.1~km.s$^{-1}$, C-MM3 is located near the center of 
gravity of the collapsing NGC~2264-C clump. An upper limit to the mass 
accretion rate onto the protostellar object C-MM3 is provided by the total 
mass inflow rate $\dot{M}_{inf} \sim$ 3$\times$10$^{-3}$ M$_{\odot}$.yr$^{-1}$
derived in \S ~6 for the central part of the NGC~2264-C.
A lower limit can be estimated from the mass accretion rate onto the less 
extreme object D-MM1 in NGC~2264-D,  
$\dot{M}_{DMM1} \sim$ 1.1$\times$10$^{-4}$ M$_{\odot}$.yr$^{-1}$ (cf \S~7).
The mass accretion rate onto C-MM3 is thus likely in the 
range 1.1$\times$10$^{-4} < \dot{M}_{CMM3} <$ 3$\times$10$^{-3}$ M$_{\odot}$.yr$^{-1}$, which is sufficient to overcome the
radiation pressure of a $\sim 10$~M$_{\odot}$ star (Wolfire \& Cassinelli 1987,
Jijina \& Adams 1996).\\
Assuming a typical core star formation efficiency $\epsilon \sim 50\% $, 
consistent with calculations of the feedback effect due the protostellar
outflow (e.g. Matzner \& McKee 2000), C-MM3, which is thought to be a $\sim$ 40~M$_{\odot}$ core, could then be the progenitor to a  $\sim 20\, M_\odot $ star. 

In the context of the turbulent core model, the formation of a 20~M$_{\odot}$
star requires a velocity dispersion at the surface of the initial prestellar 
core $\sigma_s \sim$ 1.0 km.s$^{-1}$, assuming the fiducial values of MT03 
and a clump surface density $\Sigma_{cl} \sim$~0.6~g.cm$^{-2}$. 
The model then predicts a mass accretion rate 
$\dot{m}_\star \sim$~2.0$\times$10$^{-4}$~M$_{\odot}$.yr$^{-1}$ when half 
of the final stellar mass has been accreted (i.e., end of the Class~0 phase), 
rising to a maximum value of $\sim$~2.8$\times$10$^{-4}$~M$_{\odot}$.yr$^{-1}$
at the end of the accretion phase.
In this model, the total star formation time 
from the birth of the protostar to the moment it reaches its final mass,
$m_{\star f} = 20$~M$_{\odot}$, is t$_{*f}~\sim$~1.7$\times$10$^5$ yr.
We conclude that the observed properties of C-MM3 in NGC~2264-C are  
roughly consistent with, albeit somewhat more extreme/dynamic than, 
the MT03 model. 

The dynamical time estimated in \S~6, corresponding to the time for 
the two Class~0 protostars C-MM2 and C-MM4 to reach the position of C-MM3,
is $\sim 1.7\times10^5$~yr. This timescale is similar to the formation time of a 20~M$_{\odot}$ star in the MT03 model and to the duration of
the Class~I protostellar accretion phase for low-mass stars
(e.g. Greene et al. 1994).
Therefore, significant evolution evolution of C-MM2 and C-MM4 is likely to occur before a merger can take place in the center of NGC 2264-C. Nevertheless, a strong dynamical interaction between C-MM2 and C-MM4
seems \textit{possible} before the associated protostars have achieved their final masses. Accordingly, we suggest that the formation of
an ultra-dense protostellar core through the merger of two or more
intermediate-mass {\it protostars} in the central part of a protocluster
may be a plausible route to high-mass star formation. Although in itself 
such a merger does not directly overcome the problems arising from radiation
pressure, it provides a dynamical way to build up a massive protostellar core 
with physical characteristics similar to the McKee \& Tan model. 
Albeit reminiscent of the Bonnell et al. (1998, 2002) scenario, this 
suggested route differs in that it would be based on the collision 
and coalescence of protostellar dense cores in a gas-dominated environment, 
as opposed to stellar mergers in a stellar-dominated potential.

%
%

\subsection{Conclusions: Massive star formation in clusters}

The picture of massive, clustered star formation which emerges from the present 
study is intermediate between the highly dynamic scenario of Bonnell et 
al. (1998) and the turbulent core scenario of MT03.
In agreement with the first scenario and at variance 
with the second scenario, our observations support the view that protoclusters
are in a state of global, dynamical collapse rather than approximate 
hydrostatic equilibrium. On the other hand, in contrast to the first
scenario and better agreement with the second scenario, most stars in a
cluster, including massive stars, appear to acquire their final masses while
still embedded in a gas-dominated environment, with processes occurring 
later on playing relatively little role.\\
Based on our observations, we propose a mixed scenario of protocluster 
formation according to which a pre-existing massive, turbulent clump is 
strongly compressed and induced to collapse from outside. 
Such a clump rapidly enters a phase of global, near free-fall collapse leading 
to an averaged density profile approaching $\rho \propto r^{-1.5}$ in the 
central regions. (Note that the presence of large-scale, supersonic inward
motions may well be a generic feature of {\it all} embedded protoclusters
in high-mass star-forming complexes -- Motte et al. 2005.) 
As the clump is turbulent and contains many Jeans masses 
at the onset of dynamical collapse, it quickly fragments and produces a 
number of relatively massive `cores'. These cores 
which are denser than the clump material surrounding them are collapsing on 
significantly
shorter timescales than the entire protocluster clump. Because of the high 
turbulent/dynamic ambient pressure, the cores are characterized by large 
densities, reminiscent of the MT03 model, and form individual protostars 
with large mass infall rates.
Although most of these cores have masses exceeding the local Jeans mass, 
we speculate that further fragmentation will be largely inhibited as 
the system is already dominated by strongly
converging collapse motions as opposed to random turbulent 
motions\footnote{Strictly speaking, the observations discussed here 
mostly constrain the motions of the parent clump on scales larger than 
$\sim 10^4$~AU. Higher-resolution observations with the IRAM 
Plateau de Bure interferometer have been undertaken to probe the dynamics 
and degree of fragmentation of individual MMSs or `cores' on smaller scales
and will be reported in a future paper.}. 
This contrasts with the collapse of massive turbulent cores initially in 
hydrostatic equilibrium, which yields multiple low-mass fragments according 
to recent numerical simulations (Dobbs et al. 2005). 
In other words, we speculate that the state of global,
dynamical collapse of the parent protocluster clump will strongly limit 
the process of sub-fragmentation in the cores, which may help to solve  
one of the main problems of the MT03 model of massive star formation. 
In our radiative transfer model of NGC~2264-C, the 3D turbulent velocity 
dispersion $\sigma_{turb-3D}$=0.81 km.s$^{-1}$ is smaller than 
the infall velocity V$_{inf}$~=~1.3 km.s$^{-1}$ at 
$r \simlt 3.5 \times 10^4$~AU, so that coherent infall motions dominate 
over turbulent motions at small radii. While the estimated level 
of turbulence is probably still sufficient to alter the gas motions from a 
pure radial infall pattern, we thus do not expect turbulent, Jeans-type 
fragmentation to dominate in the inner $\simlt 3.5 \times 10^4$~AU region.
On the contrary, our observational results rather suggest that two or more 
cores are in the process of {\it merging} in this central region.

We conclude that the millimeter continuum sources identified in 
the central part of NGC~2264-C are the probable precursors to 
intermediate-/high-mass stars. Powerful winds will then be likely 
generated that can clear away most of the protocluster gas
and thus effectively stop the global collapse of the system.  
In this way, the 
protocluster clump NGC~2264-C will likely evolve into a {\it revealed} star cluster
similar to the neighbouring near-IR cluster discussed by Lada \& Lada (2003) 
(see Fig.~\ref{lada.ps}).\\

\section{Summary}
 
The main results of our comprehensive millimeter study of the cluster-forming clumps NGC 2264-C and NGC 2264-D are as follows:
\begin{enumerate}
\item From our dust continuum maps, we have extracted a total of 27 compact 
millimeter sources (MMSs) in NGC~2264, 12 in clump C and 15 in clump D. 
These MMSs have a typical diameter $\sim$ 0.04~pc and masses ranging 
from $\sim 2$ to 41~M$_{\odot}$. The median mass of the MMSs is 
slightly larger in NGC~2264-C ($\sim$ 10 M$_{\odot}$ versus $\sim$ 6 M$_{\odot}$ 
in NGC~2264-D), 
which is the more massive clump with a total gas mass 
$\sim 1650\, M_{\odot}$ (compared to $\sim 1310\, M_{\odot}$ for 
NGC~2264-D). In both cases, the mass contained in the MMSs amounts 
to less than 10\% of the total mass of the parent clump.
\item The MMSs of NGC~2264 exhibit broad N$_2$H$^+$(1--0) linewidths 
tracing supersonic line-of-sight velocity dispersions 
$\sigma_{line} \sim 0.7$~km.s$^{-1}$) on $\sim 0.1$~pc scales. 
This is $\sim$~twice and $\sim$~five times as large as the line-of-sight velocity dispersions 
measured on similar scales toward the dense cores of the $\rho$-Ophiuchi main cloud 
and the isolated prestellar cores of Taurus, respectively, suggesting 
the MMSs of NGC~2264 are significantly more 
turbulent than the dense cores of both $\rho$-Oph and Taurus.
\item As much as 70~$\%$ of the MMSs of NGC~2264-C are not starless but already contain 
candidate Class~0 protostellar objects traced by the presence of shocked H$_2$ jets. 
In NGC~2264-D, we have direct evidence of the protostellar nature of only 
25$\%$ of the MMSs from the close association with 2MASS and/or MSX near-/mid-IR 
source(s). This apparent difference between NGC~2264-C and NGC~2264-D 
may be an artifact due to the lack of systematic jet/outflow surveys in NGC~2264-D. 
The high percentage of candidate Class~0 objects in NGC~2264-C may result from a 
short, efficient episode of star formation,  
as expected in the case of triggered cloud collapse.
\item In NGC~2264-C, we observe widespread blue-skewed ``infall'' line profiles 
in optically thick tracers such as HCO$^+$(3--2) or CS(3--2), consistent 
with the presence of large-scale infall motions in this clump as previously suggested 
by Williams \& Garland~(2002).
\item Furthermore, our mapping in low optical depth tracers such as N$_2$H$^+$(1--0) 
or H$^{13}$CO$^+$(1--0) reveals a new velocity feature in the central part 
of NGC~2264-C, in the form of a sharp velocity discontinuity $\sim 2$~km.s$^{-1}$ 
in amplitude centered on the millimeter source C-MM3.
\item Taken together, we interpret these two velocity features as 
the signatures of the large-scale collapse of a cylinder-like, filamentary 
structure along its long axis. Radiative transfer calculations 
confirm this view and suggest an infall velocity V$_{inf}$=1.3~km.s$^{-1}$ along 
the main axis of the NGC~2264-C prolate clump, corresponding to a total inflow rate 
$\sim 3\times10^{-3}$~M$_{\odot}$.yr$^{-1}$ toward the central Class~0 object 
C-MM3. 
\item We have also found evidence for both infall and rotation motions 
toward the Class 0 object D-MM1 in the NGC~2264-D clump. Based on 
radiative transfer modelling of our HCO$^+$ and CS observations, 
we estimate an infalling velocity V$_{inf}$=0.1~km.s$^{-1}$ which, 
given the high density of D-MM1, corresponds to a mass infall rate 
$\sim 1.1\times 10^{-4}$~M$_{\odot}$.yr$^{-1}$. 
Such an infall rate is in principle high enough to overcome the radiation pressure 
of a massive $\simgt 10\, M_\odot $ star.
\item Comparison of our observations with existing scenarios  
for massive star formation suggests that we are witnessing the initial stages 
of the formation of a high-mass $\sim $~10-20~M$_{\odot}$ star in the 
central part of NGC~2264-C. Based on this comparison, we propose a picture 
of high-mass star formation in protoclusters 
intermediate between the highly dynamic scenario of Bonnell et al. (1998) 
and the turbulent core scenario of McKee \& Tan (2003). 
In this observationally-driven picture, the large-scale, dynamical collapse 
of a massive, initially unstable clump would lead to the formation of a 
turbulent, massive, and ultra-dense core with properties reminiscent of 
the McKee \& Tan model, through the gravitational merger of two or more 
intermediate-mass Class~0 protostellar cores in the central part of the clump, 
while the potential of the system is still gas-dominated. 
\end{enumerate}

\bigskip
\noindent
{\it Acknowledgements.} We would like to thank Fr\'ed\'erique Motte 
for providing us with her source extraction procedure and for 
useful discussions on protoclusters. We also acknowledge stimulating 
discussions with Patrick Hennebelle on cloud fragmentation and collapse. 
We are grateful to an anonymous referee for constructive comments which 
helped us improve the clarity of the paper. 
The observations presented in this paper were carried out with the IRAM 30m
telescope; IRAM is supported by INSU/CNRS (France), MPG (Germany) and IGN
(Spain).

\end{document}